\documentclass[fleqn,usenatbib]{mnras}

\usepackage{newtxtext,newtxmath}
\usepackage[T1]{fontenc}

\usepackage{silence} 
\usepackage{times}
\usepackage{enumitem}
\usepackage{graphicx}
\usepackage{amsmath}
\usepackage{bm}
\usepackage{multicol}
\usepackage{graphics}
\usepackage{url}
\usepackage{epsfig}
\usepackage{wrapfig}
\usepackage{datetime}
\usepackage{ulem}
\usepackage{verbatim}
\usepackage{float}
\usepackage{lastpage}
\usepackage{enumitem}
\usepackage{natbib}
\usepackage{deluxetable}

\providecommand{\mnras}{MNRAS}
\providecommand{\apj}{ApJ}
\providecommand{\apjl}{ApJ}
\providecommand{\aj}{AJ}
\providecommand{\apjs}{ApJS}
\providecommand{\aap}{A\&A}
\providecommand{\araa}{ARA\&A}
\providecommand{\nat}{Nature}
\providecommand{\pasp}{PASP}
\providecommand{\pasa}{PASA}

\providecommand{\nar}{New Astron. Rev.}

\providecommand\zap{ZAp}   
\providecommand\aaps{A\&A,S.}

\providecommand\actaa{AcA}

\def\simlt{\mathrel{\hbox{\rlap{\hbox{\lower4pt\hbox{$\sim$}}}\hbox{$<$}}}}
\def\simgt{\mathrel{\hbox{\rlap{\hbox{\lower4pt\hbox{$\sim$}}}\hbox{$>$}}}}

\def\hst{{\it HST}}

\def\I{\,\textsc{i}}

\def\Ldot{\ensuremath{\log(L/L_{\odot})}}

\def\arcsec{$^{\,\prime\prime}$}
\def\arcmin{$^{\,\prime}$}

\def\Auchp{(Auchettl et al. in prep.)}
\def\Aucht{Auchettl et al. (in prep.)}

\title[Progenitor of SN~2019yvr]{A Cool and Inflated Progenitor Candidate for the Type Ib Supernova 2019yvr at 2.6 Years Before Explosion}

\def\northwestern{1}
\def\toronto{2}
\def\carnegie{3}
\def\melbourne{4}
\def\astro{5}
\def\ucsc{6}
\def\dark{7}
\def\uiuc{8}
\def\jhu{9}
\def\stsci{{10}}

\author[Kilpatrick~et~al.]{\href{https://orcid.org/0000-0002-5740-7747
}{Charles D. Kilpatrick}$^{\northwestern}$\thanks{Email: ckilpatrick@northwestern.edu},
\href{https://orcid.org/0000-0001-7081-0082
}{Maria R. Drout}$^{\toronto,\carnegie}$,
\href{https://orcid.org/0000-0002-4449-9152
}{Katie Auchettl}$^{\melbourne,\astro,\ucsc,\dark}$,
\href{https://orcid.org/0000-0001-9494-179X
}{Georgios Dimitriadis}$^\ucsc$, \newauthor
\href{https://orcid.org/0000-0002-2445-5275
}{Ryan J. Foley}$^\ucsc$, 
\href{https://orcid.org/0000-0002-6230-0151
}{David O. Jones}$^\ucsc$, 
\href{https://orcid.org/0000-0003-4587-2366
}{Lindsay DeMarchi}$^\northwestern$, 
\href{https://orcid.org/0000-0002-4235-7337
}{K. Decker French}$^\uiuc$, 
\href{https://orcid.org/0000-0002-8526-3963
}{Christa Gall}$^\dark$, \newauthor
\href{https://orcid.org/0000-0002-4571-2306}{Jens Hjorth}$^\dark$,
\href{https://orcid.org/0000-0002-3934-2644
}{Wynn V. Jacobson-Gal\'{a}n}$^\northwestern$, 
\href{https://orcid.org/0000-0003-4768-7586
}{Raffaella Margutti}$^\northwestern$,
\href{https://orcid.org/0000-0001-6806-0673
}{Anthony L. Piro}$^\carnegie$, \newauthor
\href{https://orcid.org/0000-0003-2558-3102
}{Enrico Ramirez-Ruiz}$^{\dark,\ucsc}$,
\href{https://orcid.org/0000-0002-4410-5387
}{Armin Rest}$^{\jhu,\stsci}$,
\href{https://orcid.org/0000-0002-7559-315X
}{C\'{e}sar Rojas-Bravo}$^\ucsc$\\
$^\northwestern$Center for Interdisciplinary Exploration and Research in Astrophysics (CIERA) and Department of Physics and Astronomy, \\
Northwestern University, Evanston, IL 60208, USA \\
$^\toronto$David A. Dunlap Department of Astronomy and Astrophysics, University of Toronto, 50 St. George Street, Toronto, Ontario, M5S 3H4 Canada \\
$^\carnegie$The Observatories of the Carnegie Institution for Science, 813 Santa Barbara St., Pasadena, CA 91101, USA \\
$^\melbourne$School of Physics, The University of Melbourne, Parkville, VIC 3010, Australia \\
$^\astro$ARC Centre of Excellence for All Sky Astrophysics in 3 Dimensions (ASTRO 3D)\\
$^\ucsc$Department of Astronomy and Astrophysics, University of California, Santa Cruz, CA 95064, USA \\
$^\dark$DARK, Niels Bohr Institute, University of Copenhagen, Jagtvej 128, 2200 Copenhagen, Denmark \\
$^\uiuc$Department of Astronomy, University of Illinois, 1002 West Green St., Urbana, IL 61801, USA\\
$^\jhu$Department of Physics and Astronomy, Johns Hopkins University, 3400 North Charles Street, Baltimore, MD 21218, USA \\
$^\stsci$Space Telescope Science Institute, 3700 San Martin Drive, Baltimore, MD 21218, USA \\
}

\begin{document}
\date{Accepted 0000, Received 0000, in original form 0000}
\pagerange{\pageref*{firstpage}--\pageref*{LastPage}} \pubyear{2020}
\maketitle
\label{firstpage}

\begin{abstract}
We present {\it Hubble Space Telescope} imaging of a pre-explosion counterpart to SN~2019yvr obtained 2.6~years before its explosion as a type Ib supernova (SN~Ib).  Aligning to a post-explosion Gemini-S/GSAOI image, we demonstrate that there is a single source consistent with being the SN~2019yvr progenitor system, the second SN~Ib progenitor candidate after iPTF13bvn.  We also analyzed pre-explosion {\it Spitzer}/IRAC imaging, but we do not detect any counterparts at the SN location.  SN~2019yvr was highly reddened, and comparing its spectra and photometry to those of other, less extinguished SNe~Ib we derive $E(B-V)=0.51\substack{+0.27\\-0.16}$~mag for SN~2019yvr.  Correcting photometry of the pre-explosion source for dust reddening, we determine that this source is consistent with a $\Ldot = 5.3 \pm 0.2$ and $T_{\mathrm{eff}} = 6800\substack{+400\\-200}$~K star.  This relatively cool photospheric temperature implies a radius of 320$\substack{+30\\-50}~R_{\odot}$, much larger than expectations for SN~Ib progenitor stars with trace amounts of hydrogen but in agreement with previously identified SN~IIb progenitor systems.  The photometry of the system is also consistent with binary star models that undergo common envelope evolution, leading to a primary star hydrogen envelope mass that is mostly depleted but still seemingly in conflict with the SN~Ib classification of SN~2019yvr. SN~2019yvr had signatures of strong circumstellar interaction in late-time ($>$150~day) spectra and imaging, and so we consider eruptive mass loss and common envelope evolution scenarios that explain the SN~Ib spectroscopic class, pre-explosion counterpart, and dense circumstellar material.  We also hypothesize that the apparent inflation could be caused by a quasi-photosphere formed in an extended, low-density envelope or circumstellar matter around the primary star.

\end{abstract}


\section{INTRODUCTION}\label{sec:introduction}

Core-collapse supernovae (SNe) are the terminal explosions of stars with initial mass $>${}$8~M_{\odot}$ \citep{Burrows95}.  This aspect of massive star evolution was empirically confirmed by the discovery of the blue supergiant progenitor of SN~1987A \citep{podsialowski+93} and subsequent discovery of over two dozen SN progenitors in nearby galaxies \citep[][and references therein, with more discovered since]{Smartt15}.  The majority of these stars are red supergiant (RSG) progenitors of hydrogen-rich type II SNe (SNe~II), although several hydrogen-poor SN~IIb progenitor stars, all of which are A--K supergiants, have also been explored in the literature \citep[notably for SNe~1993J, 2008ax, 2011dh, 2013df, and 2016gkg;][]{aldering+94,crockett+08,maund+11,vandyk+14,kilpatrick17:16gkg}.

To date, there is only one confirmed example of a progenitor star to a hydrogen-stripped SN~Ib; the progenitor of iPTF13bvn in NGC~5608 was initially identified as a compact Wolf-Rayet (WR) star in pre-explosion {\it Hubble Space Telescope} ({\it HST}) imaging \citep{Cao13} and confirmed as the progenitor by its disappearance \citep{eldridge+16,Folatelli16}.  There are numerous upper limits on the progenitor systems of other SNe~Ib in the literature \citep{Eldridge13}, implying they tend to have low optical luminosities. Overall, this relative lack of pre-explosion detections has contributed to an ongoing debate on the nature of stripped-envelope SN progenitors.

The transition from hydrogen-rich type II to hydrogen-poor type IIb to hydrogen-free type Ib SNe, and finally to helium-free type Ic SNe is commonly understood as a continuum in final hydrogen (or helium) mass in the envelopes of their progenitor stars \citep{Filippenko97,dessart+11,dessart+12,yoon+12,yoon+15,dessart+15,maund2016}.  Possible mechanisms that can deplete stellar envelope mass include radiative mass loss \citep{Heger03, Crowther07, Smith14}, eruptive mass loss \citep{Langer94,Maeder00,r-r2005,Dessart10}, and mass transfer in binary systems \citep{Woosley+94,izzard2004,Fryer07,yoon+17}.  Stars with higher initial masses or metallicities are predicted to be more stripped at the time of core collapse due to their strong radiative winds \citep{Heger03}.  However, extremely high mass stars that are predicted to efficiently deplete their envelopes have more compact and thus less explodable cores, which is thought to lead to a significant fraction of failed SNe, that is, direct collapse to a black hole with no luminous transient \citep{Burrows07,Sukhbold+16,ari2020}.  In addition, the large relative fraction of stripped-envelope SNe in volume-limited surveys \citep[i.e., SNe~Ib and Ic;][]{li+11,shivvers+17,Graur17a,Graur17b} suggests they come from a progenitor channel including stars with initial masses $<$30~$M_{\odot}$ \citep{Smith+11b,Eldridge13}.

Mass transfer in a binary system is therefore an appealing alternative mechanism to strip massive star envelopes as the majority of massive stars are observed to evolve in binaries \citep{Sana12,Kiminki12}, and binary interactions can lead to a wide variety of outcomes based on mass ratio, orbital period, and the characteristics of each stellar component \citep[e.g.,][]{2020ApJ...901...44W}. In particular, Case B \citep[during helium core contraction;][]{Kippenhahn67} or Case BB \citep[after core helium exhaustion for a star with previous Case B mass transfer;][]{Delgado81} mass transfer can remove nearly all of a star's hydrogen envelope, although this process typically stops before hydrogen is completely depleted \citep{yoon+10,yoon+15,yoon+17}.  Stars with a small amount of hydrogen remaining might also swell up in the latest stages of evolution \citep{Divine65,Habets86,Gotberg17,Laplace20} and fill their Roche lobes to restart mass transfer.  If the mass transfer is non-conservative, that is some of the material is not accreted by the companion star, this scenario can lead to dense circumstellar material (CSM) in their local environments.  Thus, when the primary star explodes the SN ejecta might encounter and shock this material, producing strong thermal continuum and hydrogen and helium line emission at optical wavelengths \citep[i.e., SN~IIn and Ibn features;][]{Vanbeveren79,Claeys11,Maund16,Yoon17,Smith17,Gotberg19}.  Thus, the final envelope mass, radius, and composition of the star can result in SNe with diverse photometric and spectroscopic properties \citep{James10} ranging from type II to type IIn to type Ic-like evolution.

One prediction from this model of binary mass transfer is that there may be a continuum between SNe with type IIb and Ib-like behaviour, depending on their final hydrogen mass. \citet{dessart+12} find that progenitor stars with as little as $10^{-3}~M_{\odot}$ hydrogen envelope mass would produce a SN whose spectra exhibit broad H$\alpha$ line emission up to 10~days after maximum light \citep[although other studies find the envelope mass can be as large as 0.02--0.03~$M_{\odot}$ with no H$\alpha$ signature;][]{Hachinger12}. Stars on either edge of this dividing line are expected to vary not only in the spectroscopic evolution of their resulting SN but also their appearance in pre-explosion imaging.
Above this threshold, spectroscopic evolution should be similar to archetypal SNe~IIb such as SN~1993J \citep{Filippenko+93, Richmond+94, Woosley+94}, and the progenitor star can inflate to radii $>$400 R$_\odot$ \citep{yoon+17,Laplace20}. Indeed, the progenitor of SN~1993J was a K-type supergiant \citep{Nomoto+93,aldering+94,Fox+14}. In contrast, stars with final hydrogen-envelope masses low enough that they would be classified as a type Ib SN prior to maximum light are only expected to inflate to radii of {\it at most} $\sim$100 R$_\odot$ \citep{yoon+12,yoon+15,yoon+17,kleiser+18,Laplace20} and in many cases remain significantly smaller. This should result in hotter progenitor stars for a given luminosity.

Intriguingly, some SNe~Ib exhibit signatures of circumstellar interaction with hydrogen-rich gas weeks to months after explosion, which suggests their progenitor stars (or binary companions) recently released this material.  The best-studied example to date is SN~2014C \citep{Milisavljevic15,Margutti+16,Tinyanont16,Tinyanont19}, which was discovered in NGC~7331 at $\approx$15~Mpc, but several other stripped-envelope SNe with similar evolution have been presented in the literature \citep[e.g., SNe~2001em, 2003gk, 2004dk, 2018ijp, 2019tsf, 2019oys;][]{Chugai06,Bietenholz14,Chandra17,Mauerhan18,Pooley19,Sollerman20,Tartaglia20} as well as the initially hydrogen-free superluminous SN iPTF13ehe \citep{Yan17}.  Although non-conservative mass transfer or common envelope ejections have been proposed as the source of this material \citep{Sun20}, it is still unclear what evolutionary pathways lead to these apparently hydrogen-stripped stars or what exact mechanism causes an ejection timed only years before explosion \citep[up to $1~M_{\odot}$ of hydrogen-rich CSM for SN~2014C in][]{Margutti+16}.

Understanding how common stripped-envelope SNe with circumstellar interactions are might aid in ruling out less likely mechanisms, but constraining the exact rate is difficult as few SNe are close and bright enough to follow to late times and stripped-envelope SNe tend to be further extinguished in their host galaxies \citep{Stritzinger18}.  Some SN~Ib exhibit clear signatures of circumstellar interaction with helium-rich material at early times \citep[so-called SNe~Ibn, with narrow emission lines of helium indicative of interaction between SN ejecta and slow-moving, circumstellar helium;][]{pastorello+08,Shivvers17b}, potentially from massive, helium-rich WR stars undergoing extreme mass loss immediately before explosion \citep{smith+17}.  However, events from this class are rare and there with significant photometric and spectroscopic diversity \citep{Hosseinzadeh17}.  \citet{Margutti+16} analyzed 183 SNe~Ib and Ic with late-time radio observations and found that 10\% exhibit evidence for rebrightening consistent with SN~2014C-like evolution, implying this phenomenon may be relatively common. However, volume-limited samples with light curves beyond $100$~days of discovery \citep[when most of these interactions occur;][]{Sollerman20} are small \citep[e.g., in][]{li+11,shivvers+17}, and so there may be an observational bias preventing precise constraints on the intrinsic rate of these interactions in SNe~Ib/c.

In this paper we discuss a progenitor candidate for the SN~Ib 2019yvr discovered in NGC~4666 on UTC 2019 December 27 12:30:14 (MJD 58844.521) by the Asteroid Terrestrial impact Last Alert System \citep[ATLAS;][]{Smith19}\footnote{SN~2019yvr is also called ATLAS19benc.  We use SN~2019yvr throughout this paper for consistency with follow-up reports.}.  We present early-time light curves and spectra of SN~2019yvr demonstrating that it resembles several other SNe~Ib and is spectroscopically most similar to iPTF13bvn, albeit with much more line-of-sight extinction than most known SNe~Ib.  Although we do not present any observations beyond 35~days from discovery, we note that SN~2019yvr exhibited signatures of circumstellar interaction at $>$150~days from discovery, with evidence for relatively narrow H$\alpha$, X-ray, and radio emission at these times \Auchp.  From this information, we infer that SN~2019yvr is similar to SN~2014C, with early-time type Ib-like evolution but transitioning around 150~days to a light curve powered by shock interaction with CSM at all wavelengths.

NGC~4666 has deep {\it Hubble Space Telescope}/Wide Field Camera 3 (\hst/WFC3) imaging in F438W, F555W, F625W, and F814W bands (roughly $BV\!RI$, respectively) that covers the site of SN~2019yvr 2.6~yr before its explosion \citep{Shappee16,Foley16,Graur18}.  Compared with limits on the progenitor stars of other SNe~Ib in the literature \citep[][]{Eldridge13} as well as the detection of the progenitor star of iPTF13bvn \citep{Cao13}, these data are among the deepest pre-explosion imaging for any SN~Ib.  We compare follow-up adaptive optics-fed imaging to the pre-explosion \hst\ images and identify a single progenitor candidate with ${\rm F555W} = 25.35 \pm 0.03$~mag and ${\rm F555W} - {\rm F814W} = 1.10 \pm 0.04$~mag (AB mag; \autoref{tab:progenitor}).

Using constraints on the interstellar host extinction to SN~2019yvr inferred from photometry and spectra of the SN itself, we characterize the progenitor candidate's intrinsic spectral shape and find it is consistent with a star with $\Ldot = 5.3 \pm 0.2$ and $T_{\mathrm{eff}} = 6800\substack{+400\\-200}$~K.  This is much cooler than most SN~Ib progenitor stars are generally thought to be \citep[including the progenitor star of iPTF13bvn;][]{Cao13,Bersten+14,Folatelli16}, implying that if the counterpart is dominated by the SN~2019yvr progenitor star it must be significantly inflated compared with expectations for SN~Ib progenitor systems.  Analyzing Binary Population and Spectral Synthesis \citep[BPASS;][]{eldridge+17} stellar evolution models, we find that the \hst\ photometry could be consistent with a $19~M_{\odot}$ progenitor star with a relatively low-mass ($\approx$1.9$~M_{\odot}$), close companion star that undergoes common envelope evolution and sheds most of its hydrogen envelope. However, all of these models predict a significant residual hydrogen envelope mass and are thus in conflict with the observed type Ib spectral class of SN~2019yvr. Therefore, we hypothesize that the progenitor star may have shed its remaining hydrogen envelope through pre-SN eruptive mass ejection in the last 2.6~yr before explosion.  Otherwise, the apparently inflated radius may be caused by a much lower mass of hydrogen forming a compact quasi-photosphere in the progenitor star's circumstellar environment soon before explosion.

Throughout this paper, we assume a distance to NGC~4666 of $m-M=30.8\pm0.2$~mag ($14.4 \pm 1.3$~Mpc) derived from the light curve of the type Ia SN ASASSN-14lp also observed in this galaxy \citep{Shappee16}.  We assume a redshift to NGC~4666 of $z=0.005080$ \citep{Allison14} and Milky Way reddening $E(B-V)=0.02$~mag \citep{Schlafly11}.

\section{OBSERVATIONS}\label{sec:observations}

\subsection{High-Resolution Pre-Explosion Images of the SN~2019\lowercase{yvr} Explosion Site}\label{sec:archival}

We analyzed \hst/WFC3 imaging of NGC~4666 obtained from the Mikulski Archive for Space Telescopes\footnote{\url{https://archive.stsci.edu/hst/}}.  These data were observed over five epochs from 2017 April 21 to August 7 (Cycle 24, GO-14611, PI Graur; see \autoref{tab:progenitor}), corresponding to 980 to 872~days (2.68 to 2.39~years) before discovery of SN~2019yvr.  Using our analysis code {\tt hst123}\footnote{\url{https://github.com/charliekilpatrick/hst123}}, we downloaded every \hst\ image covering the explosion site of SN~2019yvr.  These comprised WFC3/UVIS {\tt flc} frames calibrated with the latest reference files, including corrections for bias, dark current, flat-fielding, bad pixels, and geometric distortion.  We optimally aligned each image using {\tt TweakReg} with 1000--2000 sources per frame and resulting in frame-to-frame alignment with 0.1--0.2~pix (0.005--0.010\arcsec) root-mean-square dispersion.  We then drizzled all images in each band and epoch with {\tt astrodrizzle}.  With the drizzled F555W frame as a reference, we obtained photometry in the {\tt flc} frames of every source on the same chip as the SN~2019yvr explosion site using {\tt dolphot} \citep{dolphot}.  Our {\tt dolphot} parameters followed the recommended settings for WFC3/UVIS\footnote{\url{http://americano.dolphinsim.com/dolphot/dolphotWFC3.pdf}} as described in {\tt hst123}.  We show a colour image constructed from the F814W, F555W, and F438W frames obtained on 2017 June 13 in \autoref{fig:astrometry}.

In addition, multiple epochs of {\it Spitzer}/Infrared Array Camera (IRAC) imaging of NGC~4666 were obtained from 2005 January 4 to 2014 September 25, or roughly 15.0 to 5.3~yr before discovery of SN~2019yvr.  There was a single epoch of Channel 4 (7.9$\mu$m) imaging that observed NGC~4666 (AOR 21999872; PI Rieke), but no {\it Spitzer}/IRAC observations cover NGC~4666 in Channel 3 (5.7$\mu$m).  We downloaded the basic calibrated data ({\tt cbcd}) frames and stacked them using our custom {\it Spitzer}/IRAC pipeline based on the {\tt photpipe} imaging and reduction pipeline \citep{Rest+05,Kilpatrick18:16cfr}.  The IRAC frames were stacked and regridded to a pixel scale of 0.6~arcsec~pixel$^{-1}$ using {\tt SWarp} \citep{swarp}.  We performed photometry on the stacked frames using {\tt DoPhot} \citep{schechter+93} and calibrated our data with {\it Spitzer}/IRAC instrumental response \citep[for the cold and warm missions where appropriate;][]{irac} in the stacked frames. Based on the PSF width and average sky background, the average depth of the {\it Spitzer}/IRAC images is approximately (3$\sigma$; AB mag) 24.3~mag, 24.6~mag, and  23.0~mag at 3.6, 4.5, and 7.9~$\mu$m, respectively.

\subsection{Adaptive Optics Imaging of SN~2019\lowercase{yvr}}

We observed SN~2019yvr in $H$-band on 2020 March 8, or 72~days after discovery, with the Gemini-South telescope from Cerro Pach\'{o}n, Chile and the Gemini South Adaptive Optics Imager \citep[GSAOI;][]{GSAOI}.  We used the Gemini Multi-conjugate Adaptive Optics System \citep[GeMS;][]{GeMS} with the Gemini South laser guide star system to perform adaptive optics corrections over the GSAOI field of view (85\arcsec$\times$85\arcsec) and using SN~2019yvr itself to perform tip-tilt corrections.  We alternated observations between a field covering SN~2019yvr and a relatively empty patch of sky 4\arcmin\ to the south in an on-off-on-off pattern, totaling 1005 seconds of on-source exposure time over 39 frames.
Using the GSAOI reduction tools in {\tt IRAF}\footnote{{\tt IRAF} is distributed by the National Optical Astronomy Observatory, which is operated by the Association of Universities for Research in Astronomy (AURA) under a cooperative agreement with the National Science Foundation.}, we flattened the images with a flat-field frame constructed from observations of a uniformly-illuminated screen in the same filter and instrumental setup with unilluminated frames of the same exposure time to account for bias and dark current.  We then subtracted the sky frames from our on-source frames.

\begin{figure*}
    \includegraphics[width=\textwidth]{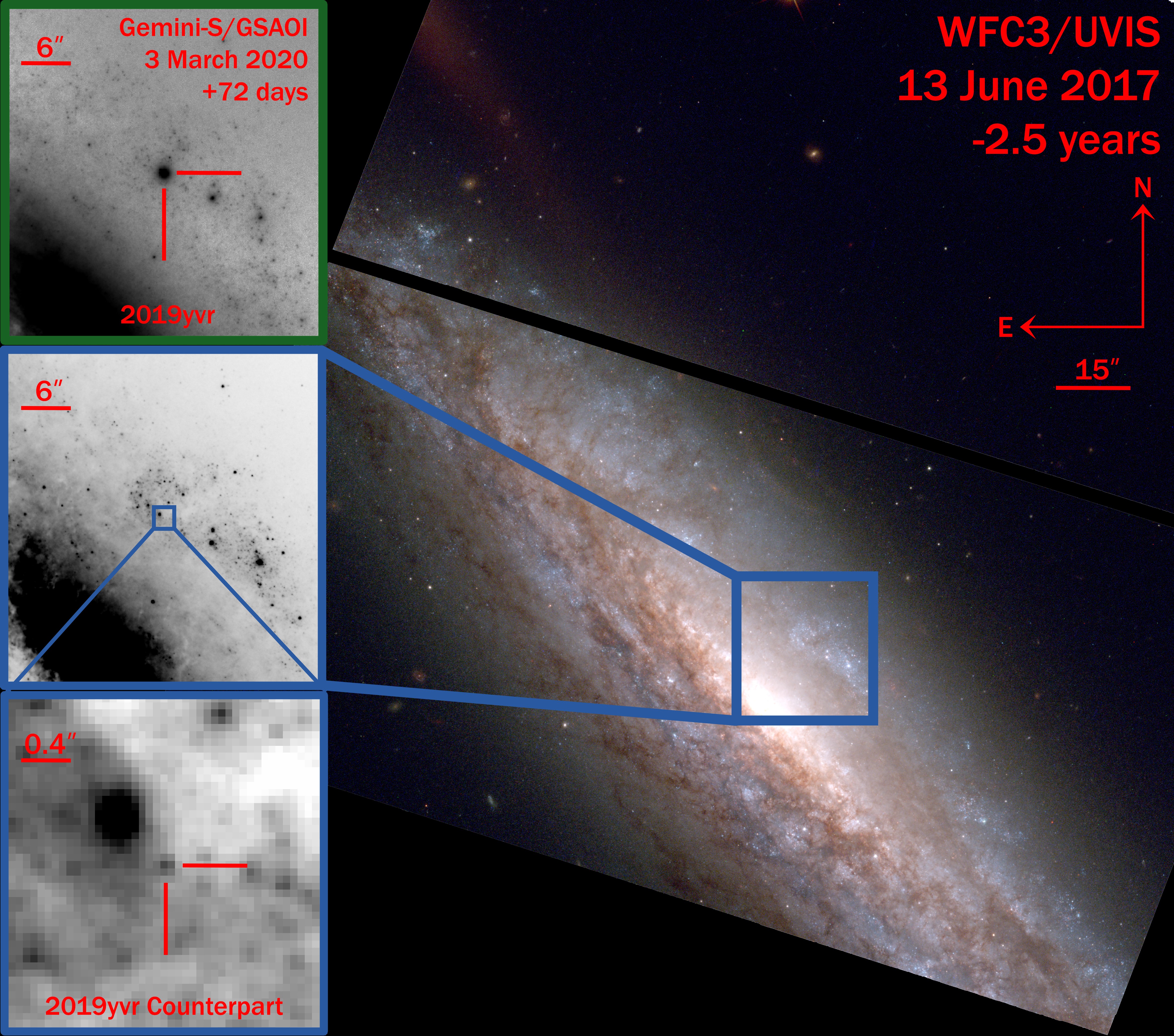}
    \caption{({\it Right}) {\it Hubble Space Telescope} imaging of the SN~2019yvr explosion site from 2.5~years before discovery consisting of F814W (red), F555W (green), and F438W (blue).  All images are oriented with north up and east to the left.  The color image on the right is 165\arcsec$\times$165\arcsec, while the left-upper and left-middle images are 38.8\arcsec$\times$38.8\arcsec, and the left-lower image is 2.4\arcsec$\times$2.4\arcsec.  The blue box denotes the approximate location of SN~2019yvr.  ({\it Upper left}): Gemini-S/GSAOI $H$-band image of SN~2019yvr obtained 67~days after discovery of the transient.  The image is centered on the location of SN~2019yvr.  ({\it Middle left}): Pre-explosion F555W imaging of NGC~4666 showing the same location as the upper left.  ({\it Lower left}): Pre-explosion F555W imaging zoomed into the blue box from the middle left.  The location of the SN~2019yvr progenitor candidate derived from our Gemini-S/GSAOI imaging is shown as red lines, which agrees with the location of a single point source as discussed in \autoref{sec:alignment}.}\label{fig:astrometry}
\end{figure*}

GSAOI has a well-understood geometric distortion pattern \citep{GSAOI-astrom}.  We used this distortion pattern to resample each on-source frame to a corrected grid, aligned the individual exposures, and constructed a mosaic from each amplifier in the on-source frames with the GSAOI tool {\tt disco-stu}\footnote{\url{http://www.gemini.edu/sciops/data/software/disco_stu.pdf}}.  Finally, we stacked the individual frames with {\tt SWarp} using an inverse-variance weighted median algorithm and scaling each image to the flux of isolated point sources observed in every on-source exposure. The final stacked frame is shown in the upper-left inset of \autoref{fig:astrometry} centered on SN~2019yvr.

\subsection{Photometry of SN~2019\lowercase{yvr}}\label{sec:imaging}

We observed SN~2019yvr with the Swope 1.0-m telescope and Direct/4K$\times$4K imager at Las Campanas Observatory, Chile from 2020 January 1 to 28 in $uBV\!gri$.  Following reduction procedures described in \citet{Kilpatrick18:16cfr}, we performed all image processing and photometry on the Swope data using {\tt photpipe} \citep{Rest+05}.  The final $BV\!gri$ photometry of SN~2019yvr were calibrated using PS1 standard sources \citep{flewelling+16} observed in the same field as SN~2019yvr and transformed into the Swope natural system following the Supercal method \citep{scolnic+15}.  In $u$-band, we calibrated our images using SkyMapper standards \citep{Onken20} in the same frame as SN~2019yvr.

We also observed SN~2019yvr with the Las Cumbres Observatory (LCO) Global Telescope Network 1-m telescopes from 2019 December 29 to 2020 February 3 with the Sinistro imagers and in $g^{\prime}r^{\prime}i^{\prime}$.  We obtained the processed images \citep[from the {\tt BANZAI} pipeline;][]{BANZAI} from the LCO archive and processed them in {\tt photpipe}, registering each image to a corrected grid with {\tt SWarp} \citep{swarp} and performing photometry on the individual frames with {\tt DoPhot} \citep{schechter+93}.  We then calibrated the $g^{\prime}r^{\prime}i^{\prime}$ photometry using $gri$ PS1 standards.

\begin{figure}
    \centering
    \includegraphics[width=0.49\textwidth]{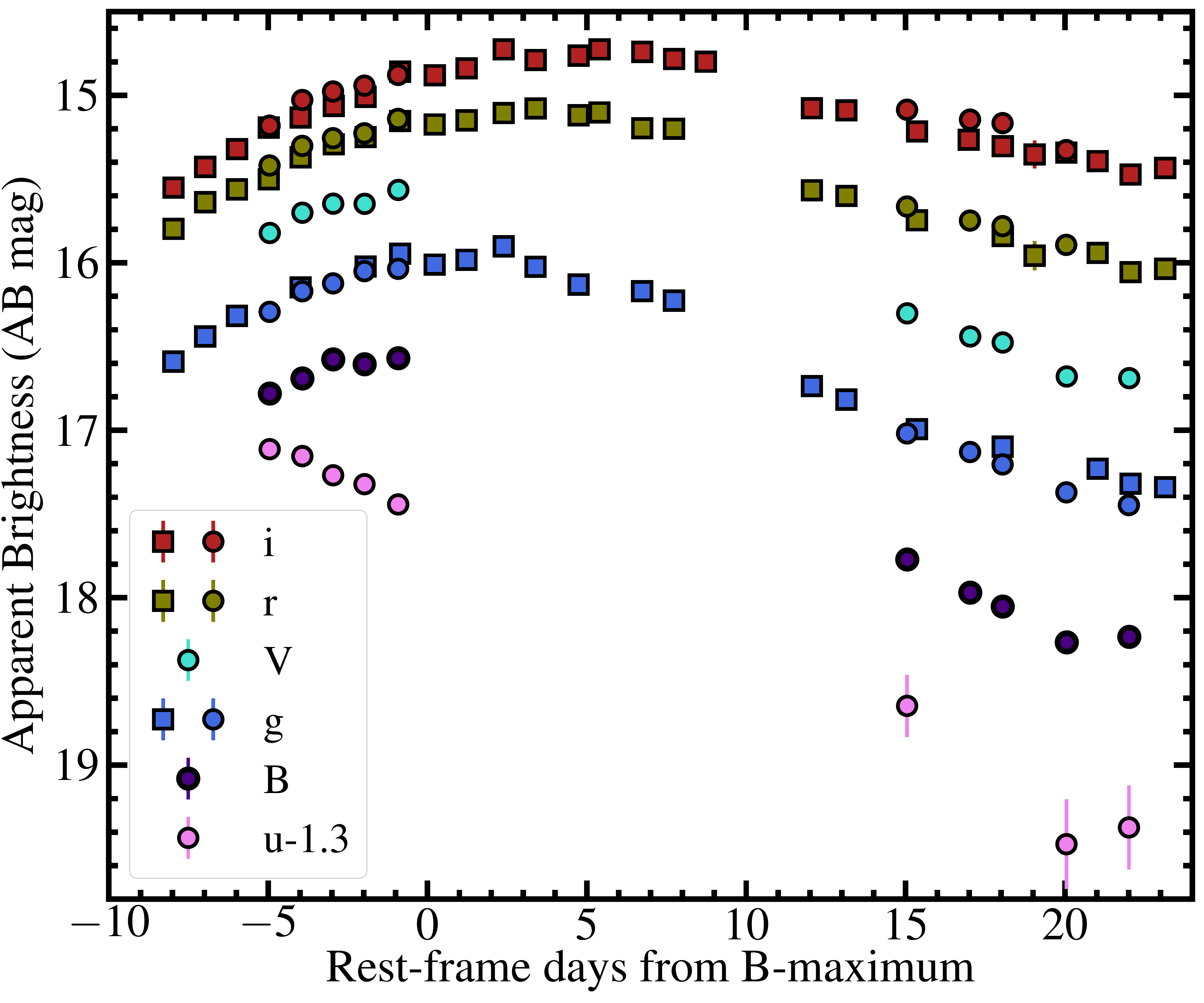}
    \caption{Swope (circle) and LCO (square) $uBgVri$ light curves of SN~2019yvr as described in \autoref{sec:imaging}.  We denote the epoch of each observation in rest-frame days (correcting for the redshift of NGC~4666 at $z=0.005080$) from $B$-band maximum light detected at MJD 58854.28 (Table~\ref{tab:photometry}).}
    \label{fig:sn-lightcurve}
\end{figure}

All Swope and LCO photometry are listed in Table~\ref{tab:photometry} and shown in \autoref{fig:sn-lightcurve}.  We estimated the time of maximum light in $V$-band by fitting a low-order polynomial to the overall light curve and derive a time of $V$-band maximum light at MJD 58853.64 (2020 January 5.64).  Detailed modelling of the light curves and inferred explosion parameters will be presented by \Aucht.

\subsection{Spectroscopy and Classification of SN~2019\lowercase{yvr}}\label{sec:spectroscopy}

\begin{figure}
    \includegraphics[width=0.49\textwidth]{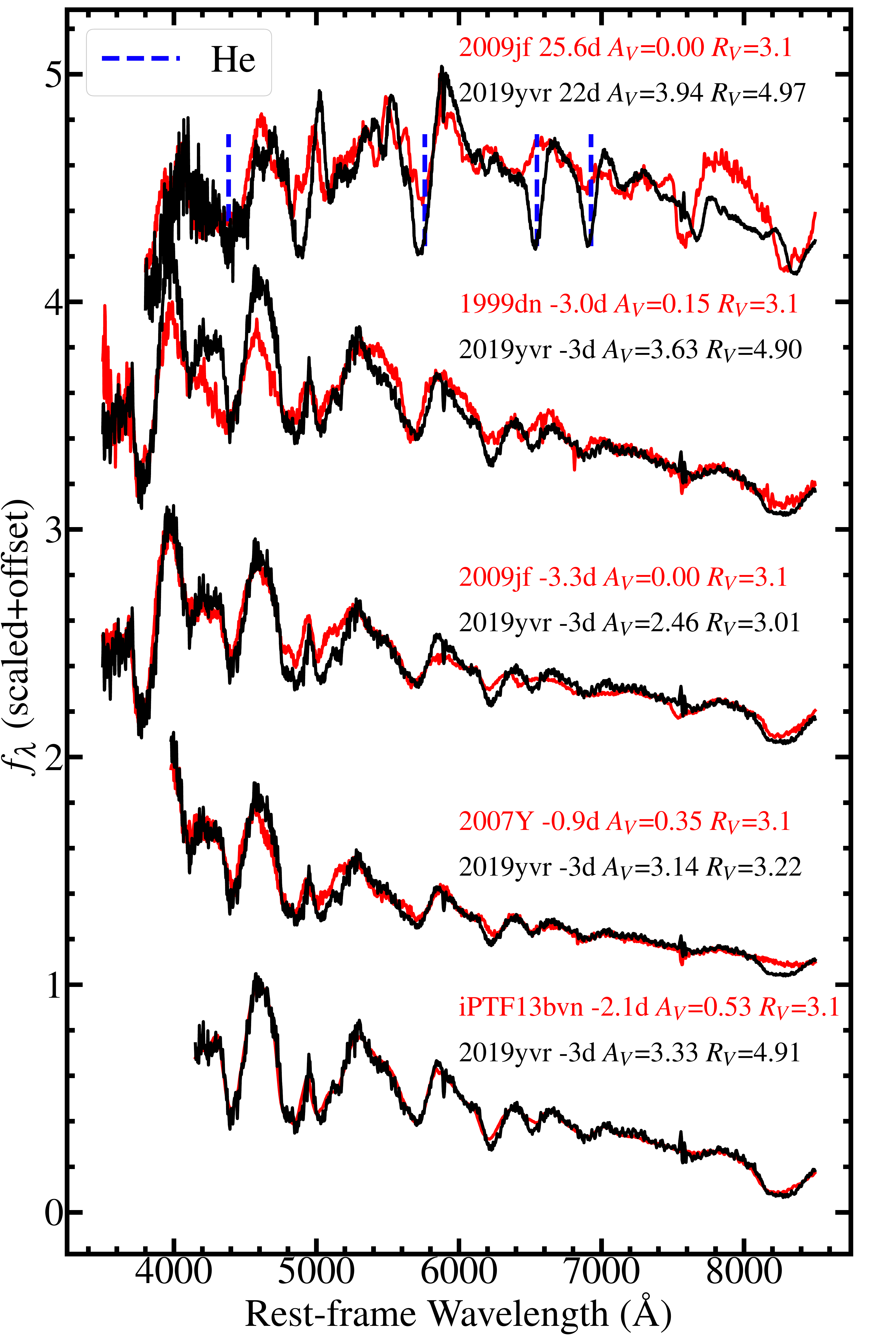}
    \caption{Our spectra of SN~2019yvr (black) with comparison to other SNe~Ib (red).  All dates are indicated with a ``d'' with respect to $V$-band maximum light.  The comparison spectra have been dereddened for Milky Way extinction based on values in \citet{Schlafly11} and dereddened for host extinction based on values in \citet[][]{Deng00,Benetti11} (for SN~1999dn), \citet[][]{Stritzinger09} (for SN~2007Y), \citet[][]{Valenti11} (for SN~2009jf), and \citet[][]{Srivastav14} (for iPTF13bvn).  We removed the recessional velocity for $z=0.005080$ from the SN~2019yvr spectra and dereddened them following the methods given in \autoref{sec:extfit}.  The best-fitting extinction and $R_{V}$ parameter are given next to each SN~2019yvr spectrum.  We highlight lines of He\I\ at $\lambda\lambda$4471, 5876, 6678, and 7065, which are present in both epochs, demonstrating that SN~2019yvr is a SN~Ib.}\label{fig:spectra}
\end{figure}

We triggered spectroscopic observations of SN~2019yvr on the Faulkes-North 2-m telescope at Haleakal\={a}, Hawaii with the FLOYDS spectrograph (Program NOAO2020A-008, PI Kilpatrick).  The spectrum was observed on 2020 January 2 roughly 5~days after the initial discovery report from ATLAS and 3~days before SN~2019yvr reached $V$-band maximum.  The observation was a 1500-s exposure at an average airmass of 1.35 and under near-photometric observing conditions.  We reduced the spectrum following standard procedures in {\tt IRAF}, including corrections for telluric absorption and correcting the wavelength solution for atmospheric diffraction using the sky lines.  The final reduced spectrum is shown in \autoref{fig:spectra}.

We also observed SN~2019yvr on the Keck-I 10-m telescope on Maunakea, Hawaii with the Low-Resolution Imaging Spectrograph (LRIS; Program 2019B-U169, PI Foley) on 27 Jan 2020, approximately 22~days after $V$-band maximum as seen from our light curve.  The observation was a 180-s exposure obtained during morning twilight at an average airmass of 1.16 and under near-photometric conditions.  We reduced these data using a custom {\tt pyraf}-based LRIS pipeline \citep{Siebert20}\footnote{\url{https://github.com/msiebert1/UCSC_spectral_pipeline}}, which accounts for bias-subtraction, flat-fielding, amplifier crosstalk, background and sky subtraction, telluric corrections using a standard observed on the same night and at a similar airmass, and order combination.  The final combined spectrum is shown in \autoref{fig:spectra}.

Our spectra reveal characteristic SN~Ib features with strong, broad absorption lines of He\I\ $\lambda\lambda$4471, 5876, 6678, and 7065 (\autoref{fig:spectra}).  These features and the lack of any apparent Balmer line emission indicate that SN~2019yvr is a typical SN~Ib, and our comparisons to other SNe~Ib such as iPTF13bvn \citep{Srivastav14} suggest it is well matched to this spectroscopic class as a whole.  From the spectrum obtained at 3 days before maximum light, we infer a velocity from Ca absorption of 22,000~km~s$^{-1}$.  We also note prominent lines of Na\I~D absorption at the redshift of NGC~4666 ($z=0.005080$).  The complete spectroscopic evolution of SN~2019yvr will be addressed by \Aucht.

\section{Extinction Toward SN~2019\lowercase{yvr} and Its Progenitor System}\label{sec:extinction}

Stripped-envelope SNe~Ib are known to occur in regions of high extinction in their host galaxies \citep{Drout11,Galbany16a,Galbany16b,Stritzinger18}.  However, if there is significant extinction due to dust in the circumstellar environment of SN~2019yvr, it may be variable between the time the \hst\ images and imaging and spectra of SN~2019yvr were obtained.  Moreover, we have no a priori constraint on the dust composition or gas-to-dust ratio in the local interstellar environment of SN~2019yvr, which is a major factor in understanding the magnitude of extinction at all optical wavelengths.

Based on the relatively low Milky Way reddening of $E(B-V)=0.02$~mag and the fact that SN~2019yvr exhibited red colours (\autoref{fig:sn-lightcurve}) and strong Na\I~D absorption, we infer that SN~2019yvr and its progenitor system are heavily extinguished by its host's interstellar and/or its own circumstellar environment.  Moreover, if we do not correct for any additional extinction, the $V$-band light curve would peak at only $-15.1$~mag. This is extremely faint compared with other SNe~Ib/c and suggests $A_{V}>1$~mag \citep[][although this inference may be affected by Malmquist bias if known samples of SNe~Ib are not representative of the overall luminosity function]{Drout11,Stritzinger18}.

Throughout the remainder of this section, we consider contextual information about the host galaxy NGC~4666, observations of SN~2019yvr, and the extinction properties of circumstellar dust around analogous stripped-envelope SN~Ib progenitor systems in order to infer the total extinction to the SN~2019yvr progenitor system.  Our goal is to derive a $V$-band extinction $A_{V}$ and reddening law parameter $R_{V}$ that can be used to estimate the total extinction in the \hst\ bandpasses as observed in pre-explosion data.

\subsection{Extinction Inferred from Na\I~D}\label{sec:naid}

One quantity that is correlated with line-of-sight reddening in both SNe \citep{Stritzinger18} and quasars \citep{Poznanski12} is the equivalent width of Na\I~D.  We detect Na\I~D in our 2019yvr LRIS spectrum with equivalent width of 4.2$\pm$0.2~\AA, which is significantly larger than the maximum Na\I~D equivalent width (2.384~\AA) from the quasars used to derive the reddening relation in \citet{Poznanski12}, implying that we might overestimate the total extinction by applying their relation.  Indeed, our measured Na\I~D equivalent width combined with the \citet{Poznanski12} relation would indicate SN~2019yvr has a light-of-sight $E(B-V)>1000$~mag, which is impossible for any extragalactic optical transient.  This finding could be due in part to saturation in the Na\I~D line for the original sample of quasars in \citet{Poznanski12}, which prevents an accurate measurement of the true column of Na\I~D as a function of the total column optical extinction.  We infer that the \citet{Poznanski12} relationship is not accurate in this high extinction and large Na\I~D equivalent width regime where we find SN~2019yvr \citep[consistent with findings in][]{Stritzinger18}.

If we instead use the relation between $A_{V}$ and Na\I~D equivalent width in \citet{Stritzinger18}, which was derived specifically from SN~Ib/c colour curves, we find SN~2019yvr has a line-of-sight extinction $A_{V} = 3.4 \pm 0.6$~mag. However, we emphasize that the validity of this relationship at such large equivalent widths has not been tested, and, more broadly, there is significant scatter in the correlation between Na\I~D equivalent width and optical extinction \citep{Phillips13}.  Therefore, we turn to other extinction indicators to better estimate the line-of-sight extinction.

\subsection{Extinction Inferred from SN~2019\lowercase{yvr} Spectra}\label{sec:ext-spectra}

Spectra and light curves of SNe~Ib similar to SN~2019yvr can be used to constrain its line-of-sight extinction.  As host extinction is a dominant systematic uncertainty in estimating intrinsic stripped-envelope SN colours, any differences in broadband colours between SNe at similar epochs can be attributed to extinction.  Here we compare our SN~2019yvr spectra to those of other SNe~Ib applying a \citet{cardelli+89} extinction law with variable $E(B-V)$ and $R_{V}$ to deredden our SN~2019yvr until they closely match.

Our template spectra are chosen from those of well-observed SNe~Ib with low host reddening ($E(B-V)<0.2$~mag) measured and reported in the literature.  These include SN~1999dn \citep[$E(B-V)=0.05$~mag;][]{Deng00,Benetti11}, SN~2007Y \citep[$E(B-V)=0.11$~mag;][]{Stritzinger09}, SN~2009jf \citep[$E(B-V)\approx0.0$~mag;][]{Valenti11}, and iPTF13bvn \citep[$E(B-V)=0.17$~mag;][]{Srivastav14}. We use spectra obtained from the Open Supernova Catalog\footnote{\url{sne.space}} \citep{OSC}.  All template spectra were chosen to correspond to roughly the same epoch relative to $V$-band maximum as one of our two SN~2019yvr spectra.  For the purposes of our fitting procedure, we assume that these extinction values are exact with no additional uncertainty.  Furthermore, we assume all template spectra experienced Milky Way-like host reddening with $R_{V}=3.1$ as $R_{V}$ is either unconstrained or poorly constrained for all of these objects.  We acknowledge that this possibly biases our $A_{V}$ and $R_{V}$ estimates for SN~2019yvr based on the spectroscopic fitting method, although $E(B-V)$ is small for our templates and so this may not be a major systematic uncertainty.
For both the SN~2019yvr and template spectra, we estimate the uncertainty in the specific flux ($\sigma_{\lambda}$) by taking

\begin{equation}
    \sigma_{\lambda} = \tilde{f}_{\lambda} \sqrt{\left | 1-\frac{f_{\lambda}}{\tilde{f}_{\lambda}} \right |},
\end{equation}

\noindent where $\tilde{f}_{\lambda}$ is the specific flux $f_{\lambda}$ passed through a smoothing function with a 50~\AA\ window and rebinned to 1~\AA\ resolution over the maximum overlap range between the SN~2019yvr and template spectrum.  Thus, the SN~2019yvr and template spectrum flux uncertainties are propagated through our entire analysis.  We then fit our LRIS and FLOYDS spectra of SN~2019yvr to the templates by calculating a dereddened spectral template $\tilde{f}_{\lambda,d}$ assuming the appropriate Milky Way extinction and the interstellar host extinction given above.  We also deredden SN~2019yvr for Milky Way extinction following the same procedure yielding $f_{\lambda,\mathrm{19yvr}}$.  For both SN~2019yvr and the template, we rescale the uncertainty $\sigma_{\lambda}$ by the same factor as the dereddened spectrum.  Finally, we derive the best-fitting host extinction $A_{V,\mathrm{19yvr}}$ and reddening law parameter $R_{V,\mathrm{19yvr}}$ by calculating $A_{\lambda}$ from \citet{cardelli+89} and minimizing the reduced $\chi^{2}$ value

\begin{equation}
    \chi^{2} = \sum^{N}_{\lambda} \frac{(\tilde{f}_{\lambda,d} - C f_{\lambda,\mathrm{19yvr}} 10^{0.4 A_{\lambda}})^{2}}{N (\sigma_{\lambda}^{2} + \sigma_{\lambda,\mathrm{19yvr}}^{2})}
\end{equation}

\noindent where $N$ is the total number of 1~\AA\ wavelength bins and $C$ is a scaling constant between the two spectra.  Thus, our spectral fitting method is primarily sensitive to the overall shape of the two spectra rather than the ratio between their fluxes.  We show our best-fitting dereddened SN~2019yvr spectra in \autoref{fig:spectra} and we list our best-fitting $A_{V}$ and $R_{V}$ parameters for SN~2019yvr in \autoref{tab:spectral-fit}.

\begin{table}
    \centering
    \scriptsize
    \begin{tabular}{c|c|c|c|c|c}
     \hline Epoch  & Template (Epoch) & $A_{V,Temp.}$ & $A_{V,\mathrm{19yvr}}$ & $R_{V,\mathrm{19yvr}}$ & $\chi^{2}$ \\
            (days) & (days)               & (mag)           & (mag)         &             & \\
     \hline\hline
            $-$3  & iPTF13bvn ($-$2.1)   & 0.53          & 3.33$\pm$0.24& 4.91$\pm$0.37 & 1.00 \\
            $-$3  & SN~2007Y  ($-$0.9)   & 0.35          & 3.14$\pm$0.29& 3.22$\pm$0.41 & 4.40 \\
            $-$3  & SN~2009jf ($-$3.3)   & 0.00          & 2.46$\pm$0.32& 3.01$\pm$0.40 & 4.08 \\
            $-$3  & SN~1999dn ($-$3.0)   & 0.15          & 3.62$\pm$0.24& 4.90$\pm$0.49 & 5.31 \\
            $+$22 & SN~2009jf ($+$25.6)  & 0.00          & 3.94$\pm$0.38& 4.97$\pm$0.56 & 17.42 \\ \hline
            Mean   &              &               & 3.3$\pm0.4$  & 4.1$\pm$0.9   &       \\ \hline

    \end{tabular}
    \caption{Our best-fitting parameters for $V$-band extinction ($A_{V}$) and $R_{V}$ inferred for SN~2019yvr based on matching to template spectra as shown in \autoref{fig:spectra} and described in \autoref{sec:ext-spectra}.  As in \autoref{fig:spectra}, the epoch of each SN~2019yvr and template spectrum is given in days with respect to $V$-band maximum light.  $\chi^{2}$ is given in units of reduced $\chi^{2}/\chi^{2}_{\mathrm{min}}$.  We give parameters for each template spectrum used and the inverse $\chi^{2}$-weighted average for $A_{V}$ and $R_{V}$.  However, see caveats in \autoref{sec:ext-spectra}.}
    \label{tab:spectral-fit}
\end{table}

\begin{figure}
    \centering
    \includegraphics[width=0.49\textwidth]{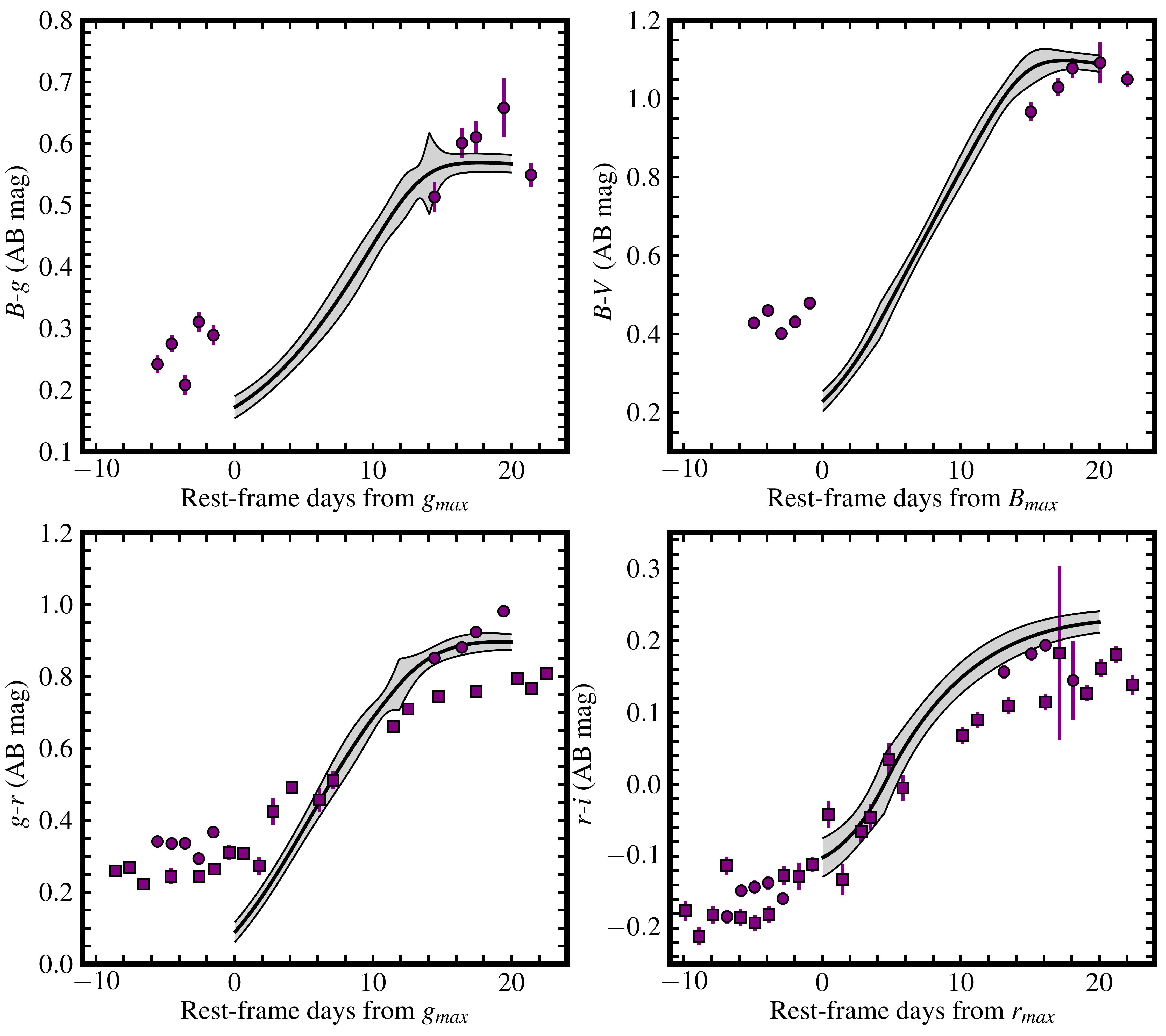}
    \caption{Colour curves of SN~2019yvr corrected for Milky Way and interstellar host extinction (with $A_{V}=2.3$~mag and $R_{V}=4.4$) as discussed in \autoref{sec:extfit}.  Circles correspond to our Swope photometry while squares are for LCO photometry.  We overplot templates for extinction-corrected SN~Ib colour curves from \citet{Stritzinger18} as black lines with the 1$\sigma$ uncertainties in each template as a gray shaded region.}
    \label{fig:colour-curves}
\end{figure}

\begin{figure}
    \centering
    \includegraphics[width=0.49\textwidth]{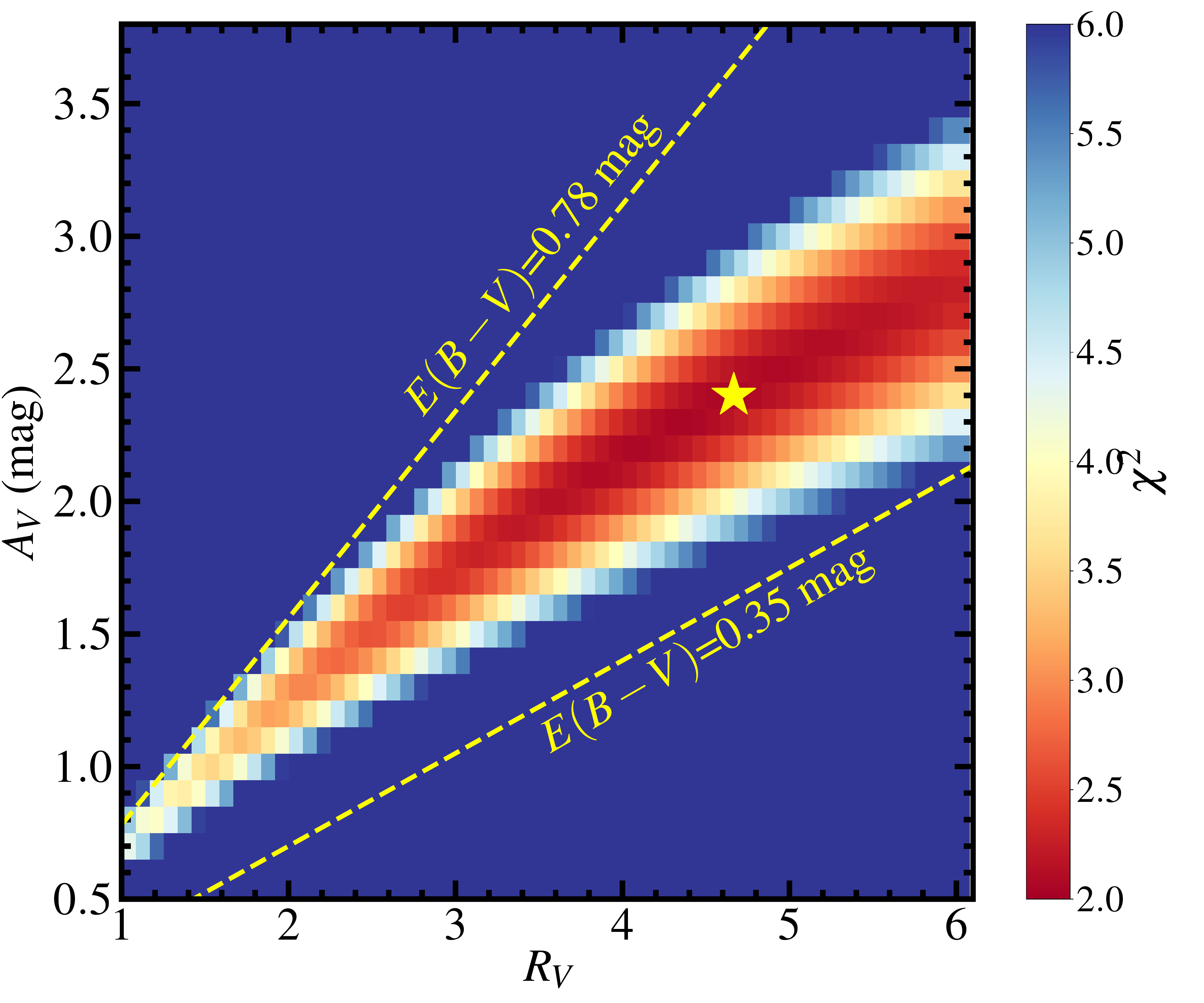}
    \caption{$\chi^{2}$ values as a function of assumed interstellar $V$-band extinction $A_{V}$ and reddening law parameter $R_{V}$ and comparing SN~2019yvr colours to the colour templates in \autoref{sec:extfit} and \autoref{fig:colour-curves}.  The best-fitting extinction parameters are $A_{V}=2.4$~mag and $R_{V}=4.7$ (yellow star) with an implied a best-fitting $E(B-V)=0.51$~mag.  The yellow dashed lines show the 1-$\sigma$ best-fitting limits of $E(B-V)$.}
    \label{fig:chi2}
\end{figure}

As reported in \autoref{sec:spectroscopy}, our SN~2019yvr spectra correspond to approximately $-3$~days and $+22$~days relative to $V$-band maximum.  For the latter spectrum, only SN~2009jf had a spectrum sufficiently close in $V$-band epoch to perform a robust comparison between spectral shape.  Thus, while the best-fitting cases all correspond to the early-time spectrum, our second epoch serves to validate the results of this analysis. In this way, we derive a line-of-sight extinction to SN~2019yvr of $A_{V}=2.4$--$3.9$~mag, although most of our best-fitting values are around $A_{V}=3.2$--$3.6$~mag.  These values are consistent with Na\I~D, but there is significant scatter in $A_{V}$, implying that there are systematic uncertainties in our method.

\subsection{Extinction Inferred from SN~2019\lowercase{yvr} Colour Curves}\label{sec:extfit}

We further investigate the interstellar host reddening using colour curve templates from \citet{Stritzinger18} and compare to our Swope and LCO colour curves of SN~2019yvr (\autoref{fig:colour-curves}).  Using a \citet{cardelli+89} reddening law, we vary the values of $A_{V}$ and $R_{V}$ in order to derive colour corrections due to interstellar reddening.  We apply these corrections to our $B-g$, $B-V$, $g-r$, and $g-i$ colour curves to find the best fit with \citet{Stritzinger18} template colours as shown in \autoref{fig:colour-curves}.  The best-fitting values are quantified with respect to the summed $\chi^{2}$ values in $B-g$, $B-V$, $g-r$, and $r-i$ and across all epochs.  The final reduced $\chi^{2}$ value for different values of $A_{V}$ and $R_{V}$ is shown in \autoref{fig:chi2}.

We find the best-fitting values using our colour curve matching are $A_{V}=2.4\substack{+0.7\\-1.1}$~mag and $R_{V}=4.7\substack{+1.3\\-3.0}$ and implying a best-fitting $E(B-V)=0.51\substack{+0.27\\-0.16}$~mag.  The value of $R_{V}$ is limited at the high end by our boundary condition that $R_{V}<6.0$. Based on well-measured values of $R_{V}$ for SN host galaxies \citep[e.g., the wide variety of SN~Ia hosts presented in][]{Amanullah15}, which tend to have $1.3 < R_{V} < 3.6$, we infer that our prior $R_{V}<6.0$ is a conservative upper bound on realistic values of the reddening law parameter.

\subsection{Final Extinction Value Adopted for SN~2019\lowercase{yvr}}\label{sec:ext-final}

Overall, the level of extinction inferred from our spectral analysis is consistent with our estimate from the $V$-band light curve as well as the value inferred from the \citet{Stritzinger18} Na\I~D relation.  While the latter relationship diverges significantly at large extinction values \citep[also similar to][]{Poznanski12}, we infer from the agreement between these three estimates that the line-of-sight host extinction inferred for SN~2019yvr is close to the value inferred from our spectral and colour curve analyses.  However, the colour curve analysis involves more independent measurements of the SN~2019yvr optical spectrum, and this analysis has been validated for several SNe~Ib by \citet{Stritzinger18}.  Thus, although there is agreement between all of our methods, we infer that $A_{V}=2.4\substack{+0.7\\-1.1}$~mag and $R_{V}=4.7\substack{+1.3\\-3.0}$ is most representative of the line-of-sight extinction to SN~2019yvr, and we use these values and our $\chi^{2}$ distribution on extinction in \autoref{fig:chi2} below.

However, the critical question to the analysis below is how much extinction did the SN~2019yvr progenitor star experience?  While it is reasonable to assume that the interstellar host extinction inferred from SN~2019yvr would be the same as the extinction that its progenitor star experienced (especially on the 2.4--15.0~yr timescale of our pre-explosion data), there could be additional sources of extinction present when the pre-explosion data were obtained to which our SN~2019yvr observations are not sensitive or vice versa.  In particular, circumstellar dust could have been present in the pre-explosion environment but vaporized soon after explosion, or else there could be material ejected by the progenitor star very soon before explosion that was not present when the {\it HST} or {\it Spitzer} data were obtained.  Below we consider both scenarios and the effects of circumstellar material and extinction on our overall data set.

\subsection{Possibility of More Circumstellar Extinction 2.6~Year Prior to Core Collapse}\label{sec:more}

All massive stars exhibit winds that pollute their environments with gas and dust \citep{Smith14}, and this material can lead to significant circumstellar extinction when the wind is dense, clumpy, and relatively cool.  Thus it is possible that the SN~2019yvr progenitor star experienced significant circumstellar extinction from a shell of dust that was vaporized before it could be observed in the SN.  \Aucht\ find evidence for a significant mass of hydrogen-rich CSM from H$\alpha$, X-ray, and radio emission.  Rebrightening in the light curve of SN~2019yvr beginning $>$150~days after discovery suggests that this material is in a shell likely at $>$1000~AU from the progenitor star, thus ejected years or decades before core collapse.

The question we address here is whether there could also be material closer to the progenitor star that contributes to circumstellar extinction but was vaporized soon after core collapse, implying that the line-of-sight extinction estimated above underestimates the extinction at the time of the {\it HST} observations.  There is no obvious sign of any such material, for example, in evolution of the Na\I~D profile or excess emission in early-time light curves and spectra.

Dust geometries and properties most likely to be associated with circumstellar extinction due to material close \citep[2--10$\times$ the photospheric radius as in][]{kochanek+12} to the progenitor star but unconstrained by our SN~2019yvr observations can be probed with our mid-infrared {\it Spitzer}/IRAC limits.  Assuming this material was present on the timescale of the IRAC observations, we model an optically thin shell of dust to our limits of 22.8, 23.1, and 21.5~mag in IRAC bands 1, 2, and 4, respectively (see \autoref{sec:alignment} for a discussion of the IRAC limits). A warm shell of gas and dust ($T>200$~K) would result in bright mid-infrared emission even in cases where it is relatively compact ($<$1~AU).  Following analysis in \citet{Kilpatrick18:16cfr}, \citet{Kilpatrick18:17eaw}, and \citet{Jacobson-Galan20}, we modeled optically thin shells of silicate dust with grain sizes $>$0.1$~\mu$m and a range of temperatures from $200$--$1500$~K.  At hotter temperatures, the dust would likely sublimate and thus would not exhibit the same extinction properties or attendant mid-infrared emission.  Similarly, a shell at large distances from its progenitor star might be so cool that it does not emit significant flux at $<10~\mu$m where our IRAC data probe, even if it has a large mass.

The dust mass limits we derive are strongly temperature dependent, with the coolest temperatures yielding the weakest limits on mass ($M_{d}<9\times10^{-4}~M_{\odot}$ and $L_{d}<4\times10^{4}~L_{\odot}$ at $200$~K) whereas hotter dust leads to relatively strong limits on dust mass ($M_{d}<2\times10^{-8}~M_{\odot}$ and $L_{d}<9\times10^{4}~L_{\odot}$ at $1500$~K).  We used the $0.1~\mu$m silicate dust grain opacities from \citet{fox+10,fox+11} to calculate these limits.  Assuming the same dust grain composition, we approximate the limits on optical depth in $V$-band as $\tau_{V} = \rho \kappa_{V} r_{\mathrm{dust}}$, where $r_{\mathrm{dust}}$ is the implied blackbody radius of the dust shell, $\rho \approx M_{d}/(4/3 \pi r_{\mathrm{dust}}^{3})$, and $\kappa_{V}$ is the opacity in $V$-band.  Under these assumptions, the optical depth must be $\tau_{V}<3$--$187$, with the strongest limits again coming from the hottest dust temperatures.

Approximating $A_{V}=0.79 \tau_{V}$ as in \citet{kochanek+12} and \citet{Kilpatrick18:17eaw}, these limits are not constraining on the total circumstellar extinction due to a compact dust shell.  Indeed, circumstellar dust absorption could be the dominant source of extinction in the SN~2019yvr progenitor system, but we would have no contextual information from the pre-explosion {\it Spitzer}/IRAC photometry to constrain the magnitude of that extinction.

The strongest argument against such a compact, warm shell of gas and dust is the lack of any hydrogen or helium emission associated with circumstellar interaction in early-time spectra or any near-infrared excess in the photometry as shown in \autoref{fig:spectra} and \autoref{fig:sn-lightcurve}.  However, these arguments are biased by the epoch of the first observations.  SN~2019yvr had a reported discovery on 2019 December 27 by ATLAS with the last previous non-detection occurring on 2019 December 11 at $>18.6$~mag in $o$-band  \citep{Smith19}.  Subsequent non-detection reports by the Zwicky Transient Facility give a more constraining non-detection in $g$-band at $>$19.5~mag on 2019 December 13\footnote{\url{https://www.wis-tns.org/object/2019yvr}}.  However, this still allows for 14~days when SN~2019yvr could have interacted with CSM in its immediate environment.  Although the first spectrum of SN~2019yvr did not exhibit evidence for flash ionization or narrow emission lines due to CSM interaction, this would not be surprising if the explosion was already more than several days old \citep[e.g., flash ionization lasted for $<$6~days for the type IIb SN 2013cu;][]{gal-yam+14}.  Deeper and higher cadence early time observations and pre-explosion limits, especially from high-resolution, near- and mid-infrared imaging, would have been needed to provide meaningful constraints on the presence and total mass of such material.

\subsection{Possibility of Less Circumstellar Extinction 2.6~Years Prior to Core Collapse}\label{sec:less}

There is strong evidence for circumstellar interaction around SN~2019yvr in optical spectra, radio, and X-ray detections starting around 150~days after discovery \Auchp.  The development of narrow Balmer lines at these late times indicates this material is hydrogen rich.  A delayed interaction points to a shell of material at a large projected separation from the progenitor ($\approx$1000~AU assuming a SN shock velocity of $\approx$10,000~km~s$^{-1}$).

A key consideration above is whether this CSM was present at the time of the {\it HST} observations or if it was ejected in the subsequent 2.6~yr before core collapse.  In the latter case, any dust synthesized in the CSM would not be present in the {\it HST} data and thus $A_{V}=2.4$~mag would be an overestimate of the extinction affecting any emission we detect in pre-explosion data.

Based on the {\it HST} observations and follow up data of the SN, we cannot constrain this scenario.  However, one prediction from this scenario would be an intermediate-luminosity transient associated with an extreme mass loss episode over this time.  We analyzed this location of the sky and found no luminous counterparts in pre-explosion imaging from the ASAS-SN Sky Patrol\footnote{\url{https://asas-sn.osu.edu/}} \citep{Shappee14,Kochanek17} or the Catalina Surveys Data Release 2 \citep{Drake09}, but these limits only extend to $<13$~mag given contamination from the bright center of NGC~4666.  Thus we cannot provide a meaningful estimate on any such CSM, but we consider the possibility that $A_{V}<2.4$~mag in our analysis of pre-explosion counterparts in Sections~\ref{sec:candidate} and \ref{sec:discussion} below.  In general, we assume that the extinction inferred from SN~2019yvr is the same between the epoch of {\it HST} observations and the time of explosion.  Overall, we consider $A_{V}=2.4\substack{+0.7\\-1.1}$~mag and $R_{V}=4.7\substack{+1.3\\-3.0}$ to represent the total line-of-sight extinction to the SN~2019yvr progenitor system at the time the pre-explosion {\it HST} imaging was obtained.

\section{The Progenitor Candidate to SN~2019\lowercase{yvr}}\label{sec:candidate}

\subsection{Aligning Adaptive Optics and Pre-Explosion Imaging}\label{sec:alignment}

We obtained positions for 114 point sources in our GSAOI adaptive optics image using {\tt sextractor} \citep{sextractor} and compared these to the positions of the same sources in the F555W {\hst}/WFC3 image as obtained in {\tt dolphot}.  From these common sources, we derived a coordinate transformation solution from GSAOI$\rightarrow$\hst.  We also derived the systematic uncertainty in this transformation by splitting our sample of common astrometric sources in half, re-deriving the coordinate transformation, and then comparing the offset between the remaining \hst\ sources and their positions from our GSAOI image and transformation.  Repeating this procedure, we are able to derive an average systematic offset between our GSAOI and \hst\ sources.  We assume the root-mean-square of these offsets dominates the error in our astrometric solution, which we find is $\sigma_{\alpha} = 0.16$ WFC3/UVIS pixels ($0.008$\arcsec) and $\sigma_{\delta} = 0.18$ WFC3/UVIS pixels ($0.009$\arcsec).

The position of SN~2019yvr in our GSAOI image corresponds to a single point source in the WFC3/UVIS imaging to a precision of $0.1$ WFC3/UVIS pixels ($\approx$0.6$\sigma$ as the uncertainty on the position of this source from our GSAOI is negligible).  We detect this source in the drizzled F555W image at 34$\sigma$ significance, and there are no other sources at the $>$5$\sigma$ level within a separation of 0.27\arcsec\ or 30 times the astrometric uncertainty. Our WFC2/UVIS photometry is listed in \autoref{tab:progenitor}.

We also examined the position of SN~2019yvr in pre-explosion {\it Spitzer}/IRAC imaging.  Using the same alignment method as above, we determined the location of SN~2019yvr in the {\it Spitzer}/IRAC stacked images using our GSAOI image of SN~2019yvr.  Our alignment uncertainty is typically $\sigma\approx0.2$ IRAC pixels ($0.12$\arcsec) from GSAOI$\rightarrow$IRAC in each channel.  We found no evidence of a counterpart in any epoch or the cumulative, stacked pre-explosion frames.  Therefore, we place an upper limit on the presence of a pre-explosion counterpart in the stacked IRAC frames by injecting and recovering artificial stars at the location of SN~2019yvr and using the native IRAC point response function for each channel.  Our pre-explosion limits for IRAC are reported in \autoref{tab:spitzer}.

\subsection{The Nature of the \hst\ Counterpart to SN~2019\lowercase{yvr}}\label{sec:cluster}

Stripped-envelope SNe are known to occur in the brightest, highest extinction, and highest metallicity regions of their host galaxies \citep{Galbany16a,Galbany16b}.  The iPTF13bvn progenitor system was identified in a relatively uncrowded region of NGC~5086 \citep{Cao13} and subsequently confirmed as the actual progenitor by its disappearance \citep{eldridge+16,Folatelli16}, but in general SNe~Ib/c are found in crowded regions of their host galaxies \citep[when the surrounding environment can be resolved, as in][]{Eldridge13}.  For example, the candidate progenitor system of the stripped-envelope SN~Ic 2017ein was in an environment with several other luminous sources \citep{Kilpatrick18:17ein}.  This fact and the counterpart's high optical luminosity suggest it may in fact have an unresolved star cluster or a chance coincidence.

The candidate progenitor system to SN~2019yvr does not appear extended in any of the WFC3/UVIS frames, with {\tt dolphot} average {\tt sharpness}=$-0.02$, {\tt roundness}=$0.36$, and classified as a bright star, which is consistent with a circular point source at WFC3/UVIS resolution.  The source is not blended with any other nearby sources and has an average {\tt crowding}=$0.09$.  Therefore, we conclude that the candidate counterpart is consistent with being a single, isolated point source in all of our images.

One possible scenario is that the candidate source is dominated by emission from multiple stars in a single system or open cluster \citep[similar to those in][]{Bastian05,Gieles06,Gieles10}.  Although the candidate source is point-like, the PSF size of {\it HST}/WFC3 in F555W is $\approx$0.067\arcsec, or 4.7~pc at the distance of NGC~4666.  Many open clusters are smaller than this, and might be so compact as to resemble a point source.  The F555W (roughly $V$-band) absolute magnitude we infer for this source is $-$7.8~mag (assuming $A_{V}=2.4$~mag), which would be extremely low luminosity for the population of clusters in \citet{Gieles06}.  Thus, while we cannot currently rule out the possibility that the source is a cluster we find it much more likely that the source is dominated by emission from a single star or star system associated with SN~2019yvr.

We estimate a single-trial probability of chance coincidence by considering that there are 3281 sources (of any type) detected at $>$5$\sigma$ in a 10\arcsec\ region surrounding the candidate SN~2019yvr counterpart in any of the \hst\ frames.  Thus, at most $6.7$~arcsec$^{2}$ or 2\% of this region is subtended by area within 3$\sigma$ (astrometric uncertainty) of any source, which is a conservative upper limit on the probability of chance coincidence between the counterpart and SN~2019yvr.  We find it is unlikely that SN~2019yvr coincides with this source by chance, although we acknowledge that this scenario cannot be ruled out definitively before we demonstrate that the source has disappeared \citep[as in the case of iPTF13bvn;][]{eldridge+16,Folatelli16}.

Given that SN~2019yvr coincides with a single, bright source, that source is point-like and isolated from nearby sources, and the relatively low likelihood of a chance coincidence, we consider this source to be a credible progenitor candidate to SN~2019yvr.  Below we assume that this object is dominated by emission from a single stellar system that hosted the SN~2019yvr progenitor star.

\begin{table}
    \centering
    \centering{WFC3/UVIS Photometry of 2019yvr Progenitor Candidate}
    \begin{tabular}{l|c|c|c|c}
\hline MJD         & Filter& Exposure (s)  & Magnitude & Uncertainty \\\hline\hline
57864.06972 & F438W & 1140 & 26.2028 & 0.2207 \\
57864.11750 & F625W & 1134 & 24.8352 & 0.0466 \\
57864.17879 & F555W & 1200 & 25.4011 & 0.0720 \\
57864.24510 & F814W & 1152 & 24.1778 & 0.0498 \\
57890.23024 & F555W & 1143 & 25.1599 & 0.0612 \\
57890.24828 & F625W & 1140 & 24.8008 & 0.0443 \\
57917.36677 & F438W & 1140 & 26.3646 & 0.4810 \\
57917.39478 & F625W & 1134 & 24.9420 & 0.0538 \\
57917.43295 & F555W & 1200 & 25.5264 & 0.0853 \\
57917.46140 & F814W & 1152 & 24.3412 & 0.0588 \\
57944.68250 & F555W & 1143 & 25.2253 & 0.0621 \\
57944.75022 & F625W & 1140 & 24.8760 & 0.0478 \\
57972.22031 & F438W & 1140 & 25.9439 & 0.2410 \\
57972.23837 & F625W & 1134 & 25.0531 & 0.0559 \\
57972.28531 & F555W & 1200 & 25.4908 & 0.0829 \\
57972.30377 & F814W & 1152 & 24.2492 & 0.0575 \\\hline
\end{tabular}
\centering{Average Photometry}
\begin{tabular}{l|c|c|c|c}\hline MJD         & Filter& Exposure (s)  & Magnitude & Uncertainty \\\hline\hline
57917.88560 & F438W & 3420 & 26.1382 & 0.1622 \\
57904.13112 & F555W & 5886 & 25.3512 & 0.0319 \\
57917.74983 & F625W & 5682 & 24.8971 & 0.0221 \\
57918.00342 & F814W & 3456 & 24.2533 & 0.0319 \\\hline
    \end{tabular}
    \caption{{\it HST} WFC3/UVIS photometry of the SN~2019yvr progenitor candidate.  All magnitudes are on the AB system.}
    \label{tab:progenitor}
\end{table}

\begin{table}
\centering{{\it Spitzer}/IRAC Pre-explosion Limits}
\begin{tabular}{l|c|c|c}\hline Average MJD & Wavelength & Exposure  & Limit \\
            & ($\mu$m)& (s) & (mag) \\\hline\hline
57917.88560 & 3.6 & 1530.0 & $>$22.8 \\
57904.13112 & 4.5 & 1864.8 & $>$23.1 \\
57918.00342 & 7.9 & 278.0  & $>$21.5 \\\hline
    \end{tabular}
    \caption{IRAC limits on the presence of a pre-explosion counterpart to SN~2019yvr progenitor candidate.  All magnitudes are on the AB system.}
    \label{tab:spitzer}
\end{table}

\subsection{Photometric Properties of the Pre-Explosion Counterpart}\label{sec:classification}

\begin{figure}
    \includegraphics[width=0.49\textwidth]{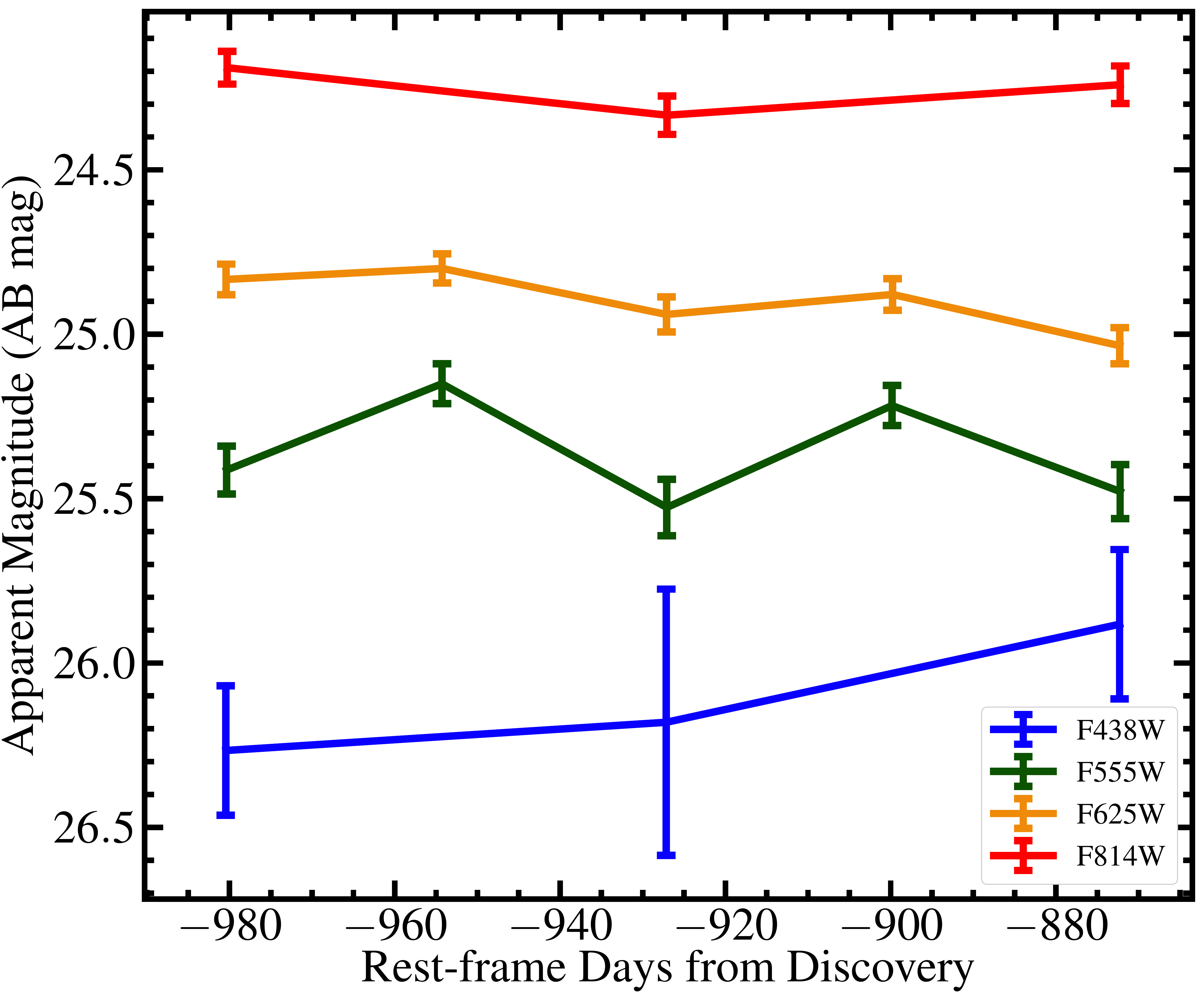}
    \caption{The pre-explosion light curve of the SN~2019yvr progenitor candidate in all four {\it HST} filters for which we have imaging.  The source is not significantly variable, with at most $0.47$~mag peak-to-peak variations as discussed in \autoref{sec:classification}.}\label{fig:lightcurve}
\end{figure}

We show the light curve of the SN~2019yvr progenitor candidate at times relative to explosion in \autoref{fig:lightcurve}.  The source is relatively stable with at most $0.47$~mag peak-to-peak variability (corresponding to 3.4$\sigma$) in F555W over a baseline of 110~days.  Thus we infer that the progenitor candidate did not exhibit any extreme variability with constant flux at the $<$0.24~mag level in all bands.  This suggests that if the counterpart is a star, it was not in an eruptive or any other highly variable phase during these observations, as these events are typically accompanied by large differences in luminosity or colour \citep[as in pre-SN outbursts associated with SNe~IIn, e.g.,][]{smith+09,mauerhan+13,Kilpatrick18:16cfr}.

Thus we are confident that the average photometry across all four {\it HST} bands in which we detect the progenitor candidate is representative of its overall spectral energy distribution (SED).  Taking an inverse-variance weighted average of across all epochs in each band, we derive average photometry $m_{\rm F438W}=26.138\pm0.162$~mag, $m_{\rm F555W}=25.351\pm0.032$~mag, $m_{\rm F625W}=26.897\pm0.022$~mag, and $m_{\rm F814W}=24.253\pm0.032$~mag as shown in \autoref{tab:progenitor}.  Temporarily, ignoring any correction due to host extinction but accounting for Milky Way extinction, the source has $m_{\rm F555W}-m_{\rm F814W}$ (roughly $V-I$) of 1.065$\pm$0.045~mag and $M_{\rm F555W}=-5.5$~mag assuming our preferred distance modulus above (both values are in AB mag).  This is roughly consistent with temperatures of $T_{\mathrm{eff}}=3360$~K, which is broadly comparable, although slightly hotter, than most terminal RSGs at solar metallicity \citep{choi+16}.  This suggests that the source either has a cool photosphere or is heavily extinguished, in agreement with expectations from our analysis of SN~2019yvr.  However, if the source is extinguished due to CSM, there is no clear evidence from pre-explosion variability whether any of this material was ejected during the window of the {\it HST} or {\it Spitzer} observations.

\subsection{Comparison to Blackbodies and Single-star Spectral Energy Distributions}\label{sec:single}

Assuming that the counterpart is dominated by the SED of a single star, we estimate the luminosity and temperature of that star by fitting various SED models to the \hst\ photometry.  Broadly, we use blackbody and stellar SEDs obtained from \citet{pickles+10}.

We use a full forward modeling and Monte Carlo Markov Chain (MCMC) approach to simulate the in-band apparent magnitudes assuming the distance above and drawing extinction ($A_{V}$) and reddening ($R_{V}$) parameters following the $\chi^{2}_{\mathrm{ext}}$ probability distribution from our light curve analysis in \autoref{sec:extinction} and as shown in \autoref{fig:chi2}. For a blackbody with a given effective temperature $T_{\mathrm{eff}}$ and luminosity $L$ as well as extinction values drawn from the $\chi^{2}$ distribution discussed above, we simulate an intrinsic, absolute magnitude $M_{i}$ in each band $i$ and convert to an apparent magnitude $m_{i}$ with in-band Milky Way extinction $A_{{\rm MW},i}$, the implied host extinction $A_{H,i}$, and our preferred distance modulus $\mu=30.8$~mag.  Thus the simulated apparent magnitude depends on both the intrinsic model parameters (i.e., $T_{\mathrm{eff}}$,$L$) and extinction parameters ($A_{V}$, $R_{V}$) drawn from $\chi^{2}_{\mathrm{ext}}$.  We then estimate the log likelihood for our MCMC ($\chi^{2}_{\mathrm{eff}}$) using the observed magnitude $m_{o,i}$ and uncertainty $\sigma_{o,i}$ from \autoref{tab:progenitor} as

\begin{equation}
    \chi^{2}_{\mathrm{eff}}=\sum_{i} \left(\frac{m_{i} - m_{o,i}}{\sigma_{i}}\right)^{2} + \chi^{2}_{\mathrm{ext}}(A_{V},R_{V}).
\end{equation}

In this way we incorporate the differences between the observed and forward-modeled magnitudes as well as between the values of $A_{V}$ and $R_{V}$ for each trial and the best-fitting values from our colour curve template fitting.  Assuming a blackbody SED, we estimate the best-fitting parameters $T_{\mathrm{eff}} = 7700\substack{+900\\-1000}$~K and $\Ldot = 5.3\substack{+0.2\\-0.3}$.  Although we include the distance modulus uncertainty in our luminosity (and radius) uncertainty estimates, we did not include this value in our fitting method as it does not affect the overall shape of the SED.  The implied photospheric radius for the best-fitting blackbodies are $R=250\pm30~R_{\odot}$.  As we incorporate $A_{V}$ and $R_{V}$ from the light curve analysis into our models, we also constrain these parameters with best-fitting values $A_{V}=2.8\substack{+0.3\\-0.4}$~mag and $R_{V}=5.2\substack{+0.8\\-0.7}$ assuming the intrinsic blackbody spectrum.

\begin{figure}
    \includegraphics[width=0.49\textwidth]{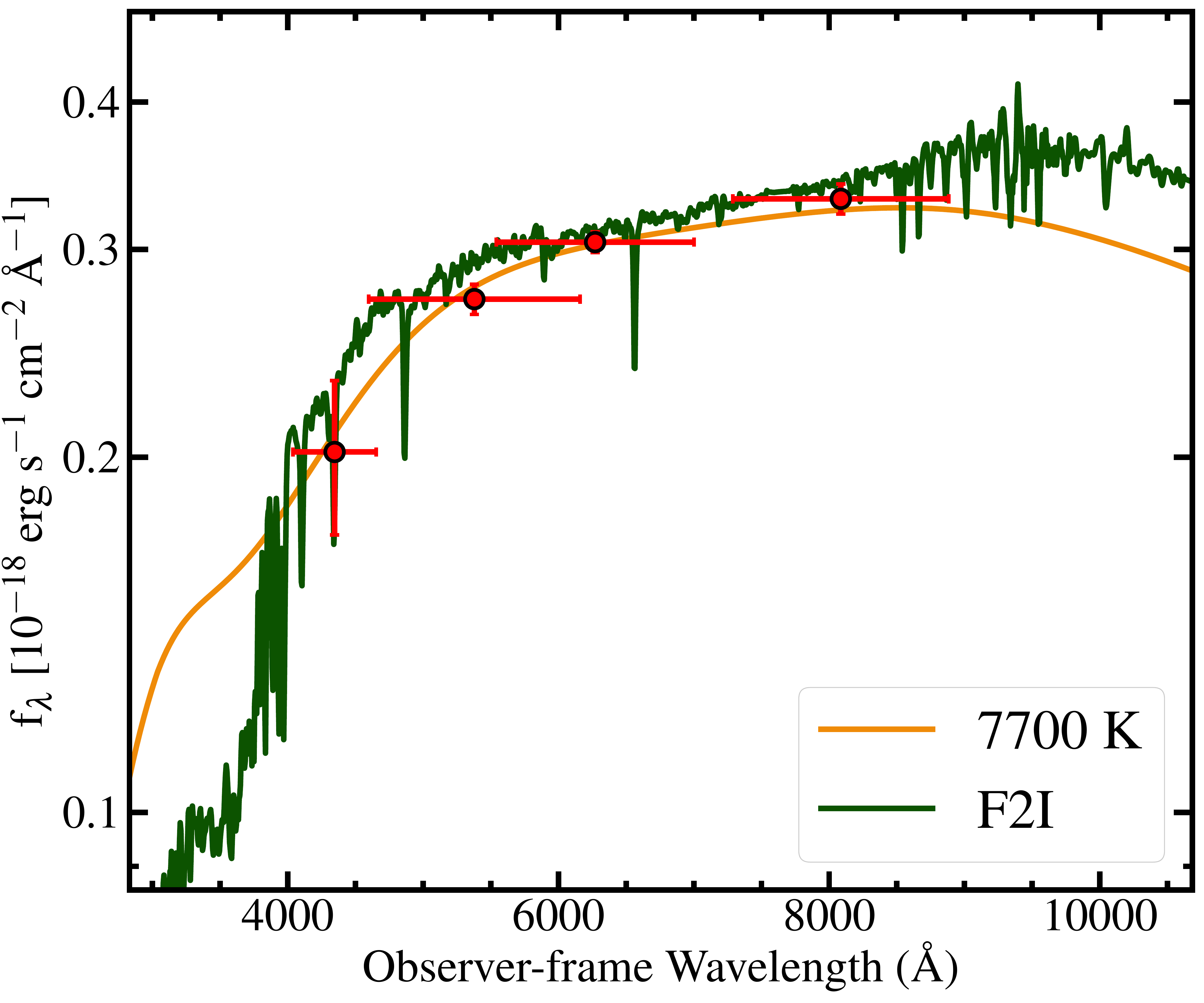}
    \caption{The best-fitting SEDs to the average pre-explosion {\it HST} photometry of the SN~2019yvr progenitor candidate.  Our best-fitting blackbody has $T_{\mathrm{eff}}=7700$~K, while the best-fitting single-star \citet{pickles+10} model is a F2I star with $T_{\mathrm{eff}}=6800$~K.  In both cases, the implied luminosity is consistent with being \Ldot$\approx$5.3.  The SEDs are completely forward-modeled in observed flux, thus they include the apparent best-fitting interstellar host and Milky Way extinction.  Although we include the distance uncertainty in our luminosity estimate, the error bars in this figure only include measurement uncertainty and uncertainty on extinction.  We simply scale the integrated flux density by our preferred distance.}\label{fig:sed}
\end{figure}

\begin{figure}
    \centering
    \includegraphics[width=0.49\textwidth]{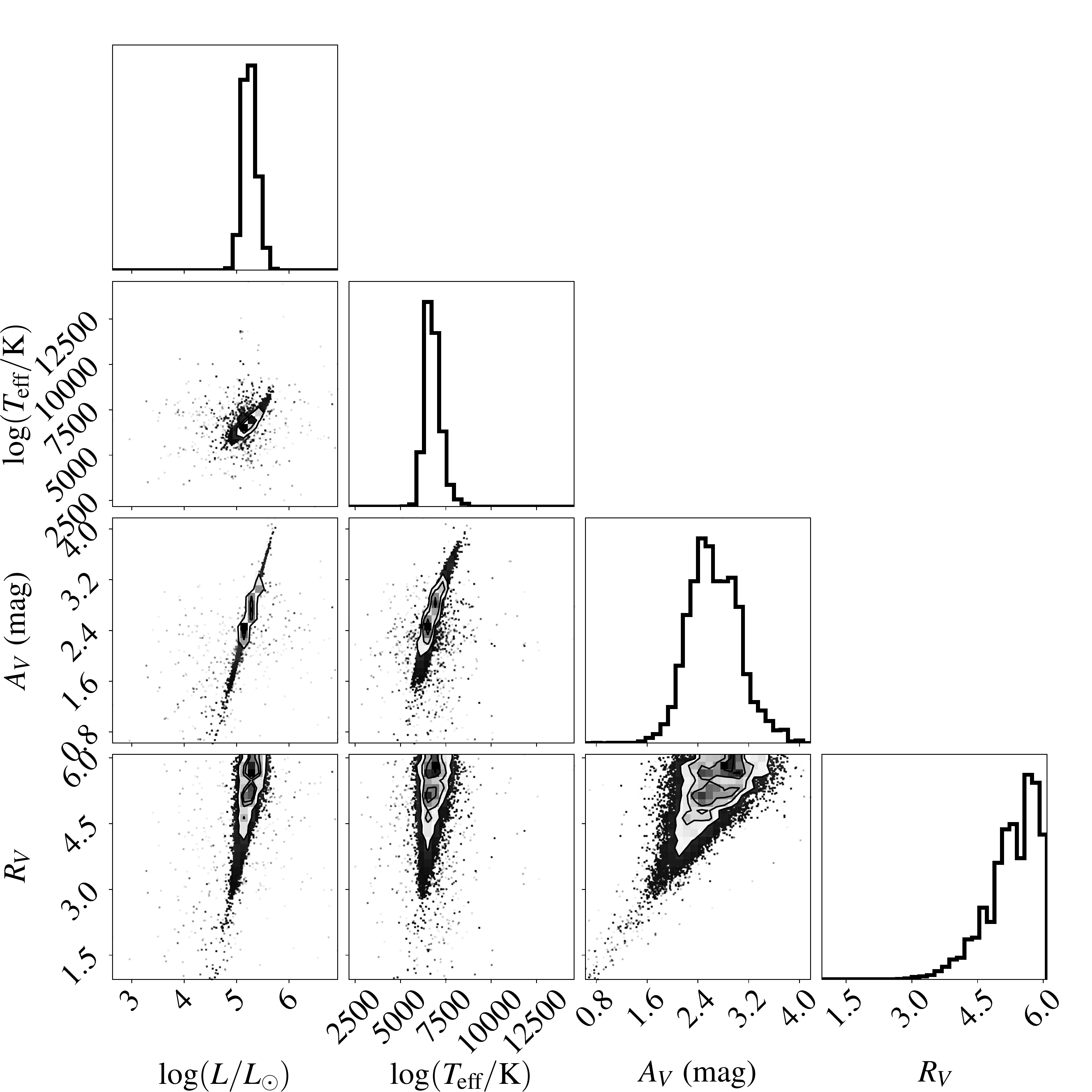}
    \caption{Corner plot showing the correlation between our fit parameters \Ldot and $\log(T/\mathrm{K})$ of a single star following a \citet{pickles+10} as well as host extinction $A_{V}$ and host reddening $R_{V}$ as described in \autoref{sec:single}.  The contours show the 1, 2, and 3$\sigma$ best-fitting values derived from all of our samples.  Although the luminosity and $V$-band host extinction are highly correlated, the resulting luminosity and temperature are tightly constrained.}
    \label{fig:corner}
\end{figure}

We also compared our photometry to single-star SEDs from \citet{pickles+10}.  We use stars of all spectral classes, fitting only to a scaled version of the stellar SED as a function of effective temperature.  Using the same MCMC method, our walkers drew a temperature randomly and then chose the stellar SED with the closest effective temperature.  The best-fitting SEDs are consistent with stars in the F4 to F0 range (intrinsic $T_{\mathrm{eff}} = 6800\substack{+400\\-200}$~K; see \autoref{fig:sed} and \autoref{fig:corner}) with an implied luminosity $\Ldot = 5.3 \pm 0.2$ and a photospheric radius $R = 320\substack{+30\\-50}$~$R_{\odot}$.  Thus the best-fitting values are broadly consistent between blackbody and \citet{pickles+10} model SEDs. There is some systematic uncertainty in the exact temperature of the latter models given the sampling of the \citet{pickles+10} spectra, which are increasingly sparse for hotter stars.  However, this effect is small at temperatures 5000--10,000~K where there are 40 spectra of varying spectral classes.

Our treatment of $A_{V}$ and $R_{V}$ is identical to the forward modeling approach for the blackbodies above, and we derive best-fitting values of $A_{V}=3.1\substack{+0.3\\-0.2}$~mag and $R_{V}=5.9\substack{+0.1\\-0.4}$.  In both the blackbody and stellar SED models, our luminosity estimates are on the high-luminosity end for observed core-collapse SN progenitor stars \citep[e.g., in][]{Smartt15} but consistent with most of the SN~IIb progenitor stars (see Figure~\ref{fig:hr}).

As a further check on the effect of extinction on our derived parameters, we show the relationship between the host extinction $A_{V}$ and reddening law parameter $R_{V}$ and the implied temperature and luminosity for the \citet{pickles+10} models in \autoref{fig:corner}.  Luminosity is highly correlated with variations in $A_{V}$, with $A_{V}<2.4$~mag implying a lower luminosity but also a significantly cooler photospheric temperature.  In general, such a cool photosphere is associated with a massive hydrogen envelope, which is in tension with the SN~Ib spectroscopic class (see Section~\ref{sec:discussion} for further discussion).  In contrast, a larger extinction value would imply a significantly higher luminosity and hotter temperature, although no combination of parameters we considered for the stellar SED fits allowed an effective temperature $>$10,000~K to within the 3-$\sigma$ level.  This is in stark contrast with the progenitor of iPTF13bvn with $T_{\mathrm{eff}}\approx45,000$~K \citep{Cao13,Bersten+14,eldridge+16,Folatelli16} and He stars generally, which tend to have effective temperatures $>$20,000~K \citep[as discussed for the progenitors of SNe~Ib in][]{yoon+15}.

Overall, our constraints on $A_{V}$ and $R_{V}$ enable a relatively tight fit temperature and luminosity as demonstrated in \autoref{fig:corner}.  The minimum $\chi^{2}/\mathrm{degrees of freedom}=1.5$, which suggests that a single, extinguished star is well matched to our data and we cannot effectively constrain scenarios with more free parameters, such as the inclusion of another star to the overall SED.  However, while this analysis might accurately reflect the SN~2019yvr progenitor star's evolutionary state at 2.6~yr before explosion, it does not place any specific constraints on the pathway that led to this configuration.  We further explore the implications of a SN~Ib progenitor star with these properties and the implications for a single-star origin in \autoref{sec:cool}.

\subsection{Comparison to Binary Star Models}\label{sec:binary}

We also compare the SED of the SN~2019yvr progenitor candidate to binary stellar evolution tracks from BPASS \citep{eldridge+17}, comprising 12,663 binary star models at a single metallicity.  BPASS provides physical parameters from the binary star system throughout its evolutionary sequence as well as in-band absolute magnitudes for the individual components and binary system as a whole.  Our analysis involved a direct comparison between the F438W, F555W, F625W, and F814W magnitudes of the SN~2019yvr counterpart and the total binary emission estimated via BPASS synthetic magnitudes in F435W, F555W, SDSS $r$, and F814W, respectively.  As BPASS magnitudes are provided in Vega mag, we transformed our AB mag photometry to Vega mag using the relative Vega mag $-$ AB mag zero points for all four WFC3/UVIS filters \citep[0.15, 0.03, $-$0.15, and $-$0.42~mag, as in][]{WFC3UVIS}.

Assuming that the SN~2019yvr progenitor is one component of a binary system, we determined what BPASS evolutionary tracks have a terminal state consistent with the SN~2019yvr candidate photometry.  Based on the gas-phase metallicity estimate of NGC~4666 in \citet{Pan+19}, which is consistent with Solar metallicity, we restrict our analysis to BPASS models with fractional metallicity $Z=0.014$.  However, we emphasize that the \citet{Pan+19} metallicity estimate is derived from spectra obtained toward the center of NGC~4666 rather than at the site of SN~2019yvr, and so the true metallicity of the SN~2019yvr progenitor system may be significantly different.  Otherwise, we consider evolutionary tracks for all BPASS initial masses ($M_{\mathrm{init}}=0.1$--$300~M_{\odot}$), mass ratios ($q=0.1$--$1.0$), and binary periods ($\log(P/\mathrm{1~day})=0$--$4$).

We used the same MCMC method as above with walkers drawing from initial masses, mass ratios, and periods and comparing the terminal absolute magnitude and colours of the closest model in parameter space to the SN~2019yvr counterpart photometry.  We also included $A_{V}$ and $R_{V}$ as free parameters, but with the walkers drawing from the same $\chi^{2}$ distribution for these parameters as in the blackbody and stellar SED fits above.  For our BPASS fits, the best-fitting models correspond to initial mass $M_{\mathrm{init}}=19\pm1~M_{\odot}$, initial mass ratio $q=0.15\pm0.05$ and initial period $\log(P/\mathrm{1~day})=0.5\pm0.2$.  We show our a Hertzpsrung-Russell diagram with the best-fitting model ($M_{\mathrm{init}}=19$, $q=0.1$, and $\log(P/\mathrm{1~day})=0.6$) in \autoref{fig:hr} and the luminosity and temperature derived from our \citet{pickles+10} stellar SED fits.

We performed our fits by comparing the observed photometry of the SN~2019yvr progenitor candidate to the apparent magnitudes inferred for the combined flux of both stars in the BPASS models\footnote{In our BPASS v2.2.1 fits, we examined columns 53--73, representing the absolute magnitude and colours from the combined flux of both the primary and companion star.}, and so we are sensitive to scenarios where the flux from either the primary or companion star dominates the total emission.  In all of the best-fitting models and all four bands we consider, the counterpart is dominated by emission from a $\approx$19$~M_{\odot}$ primary star and the companion contributes very little to the overall flux.  We found no other binary scenarios where the total flux was consistent with our photometry at the time one of the stars terminated, including scenarios where the secondary star produced the SN explosion instead (i.e., in a neutron star or black hole binary).

In the best-fitting model, the terminal state of the SN progenitor is $\Ldot = 5.3$ and $T_{\mathrm{eff}}=7300$~K with a terminal mass of $M_{\mathrm{final}}=7.3~M_{\odot}$, implying a consistent luminosity but a slightly warmer temperature than we derive from the \citet{pickles+10} models.  Similar to above, there are no models at the $<$3-$\sigma$ level where the exploding star has a terminal temperature $>$11,000~K.  In the best-fitting model, the secondary star (with an initial mass of $1.9~M_{\odot}$) is mostly unchanged with only $0.009~M_{\odot}$ of material accreted by the time the primary reaches core collapse, implying that most of the mass transfer in this model was non-conservative.  In addition, the BPASS models predict that $0.047~M_{\odot}$ (0.6\% mass fraction) of hydrogen remains in the primary in its terminal state and no model with $<${}$0.038~M_{\odot}$ is consistent with our \hst\ photometry at the 3-$\sigma$ level.  The best-fitting extinction values for this BPASS model was $A_{V}=2.6\pm0.3$~mag and $R_{V}=5.1\substack{+0.9\\-2.1}$.

The primary effect of the BPASS evolutionary models compared to single-star models is the inclusion of Roche-lobe overflow (RLOF). For our specific best-fitting model, RLOF turns on in the post-main sequence phase (i.e., Case B mass transfer; shown with a square in \autoref{fig:hr}), and continues through the end of the primary star's evolution.  In particular, mass-loss due to RLOF follows the prescription for common-envelope evolution (CEE) as the radius of the primary star is smaller than the binary separation throughout post-main sequence evolution \citep[following prescription in][]{eldridge+17}.  The binary separation is only $8.6~R_{\odot}$ starting in the post-main sequence and at the onset of CEE, and so the primary mass-loss rate increases significantly to 1--5$\times10^{-4}~M_{\odot}~\mathrm{yr}^{-1}$.  This common envelope mass loss phase largely determines the final mass and state of the primary star as it is larger than wind-driven mass loss by a factor of $\approx$1000.

In BPASS, the onset of CEE is highly correlated with small binary separations and low mass ratios for stars with $M_{\mathrm{init}}>5~M_{\odot}$ \citep{eldridge+17}.  Thus stars that terminate near the progenitor candidate in the Hertzsprung-Russell diagram require a specific mass-loss scenario where CEE can strip most of the hydrogen envelope but leave a small amount (at least $0.038~M_{\odot}$ according to our models), leading to relatively tight constraints on binary mass ratio and period for our BPASS fits.  However, these parameters are subject to significant systematic uncertainty in terms of the CEE and mass loss prescriptions assumed. In Section~\ref{sec:cool} we discuss whether these best-fit binary models to the pre-explosion photometry of SN\,2019yvr are consistent with its classification as a type Ib SN.

\section{What Progenitor Systems Could Explain SN~2019\lowercase{yvr} and the Pre-explosion Counterpart?}\label{sec:discussion}

\begin{figure}
    \centering
    \includegraphics[width=0.49\textwidth]{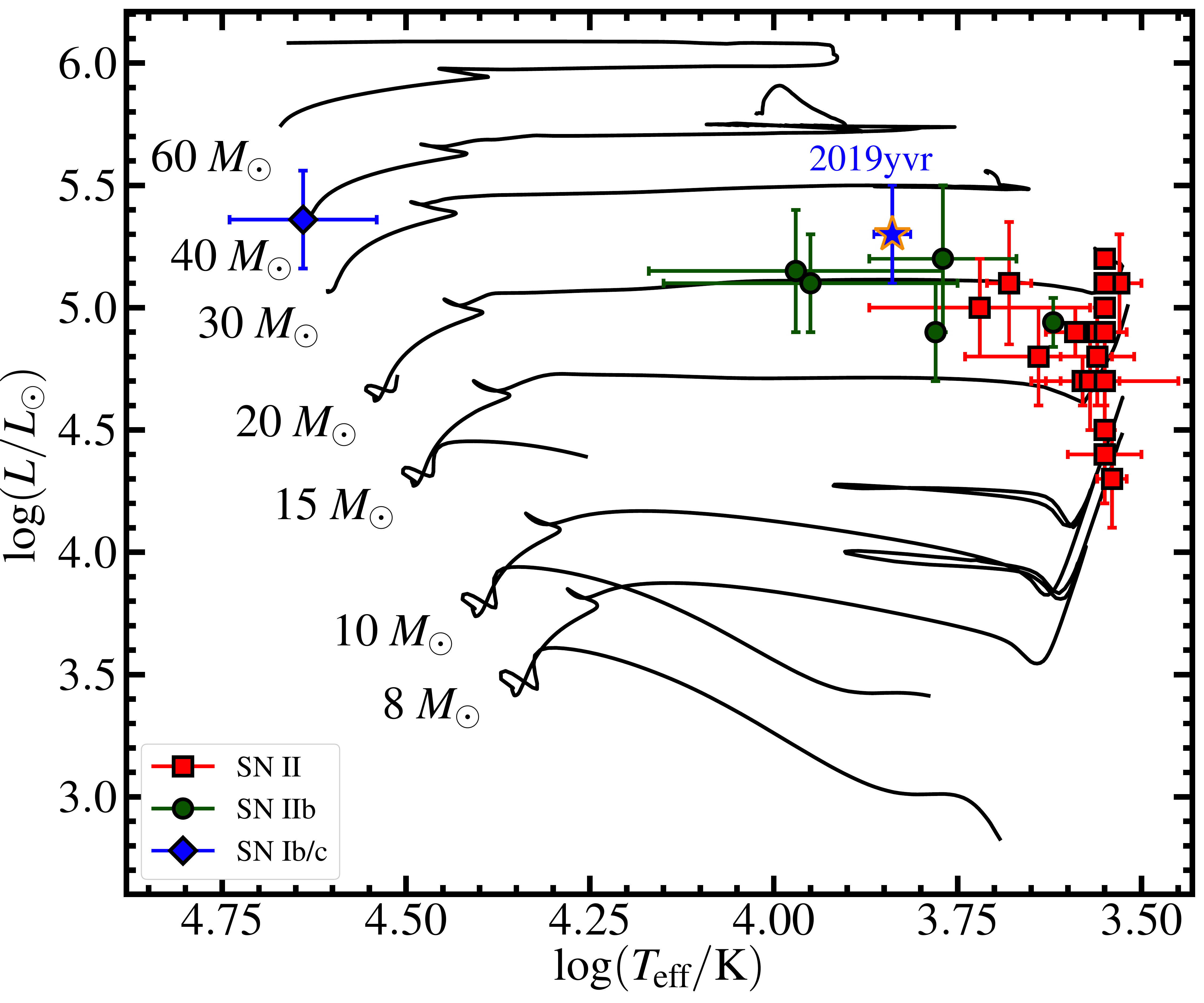}
    \includegraphics[width=0.49\textwidth]{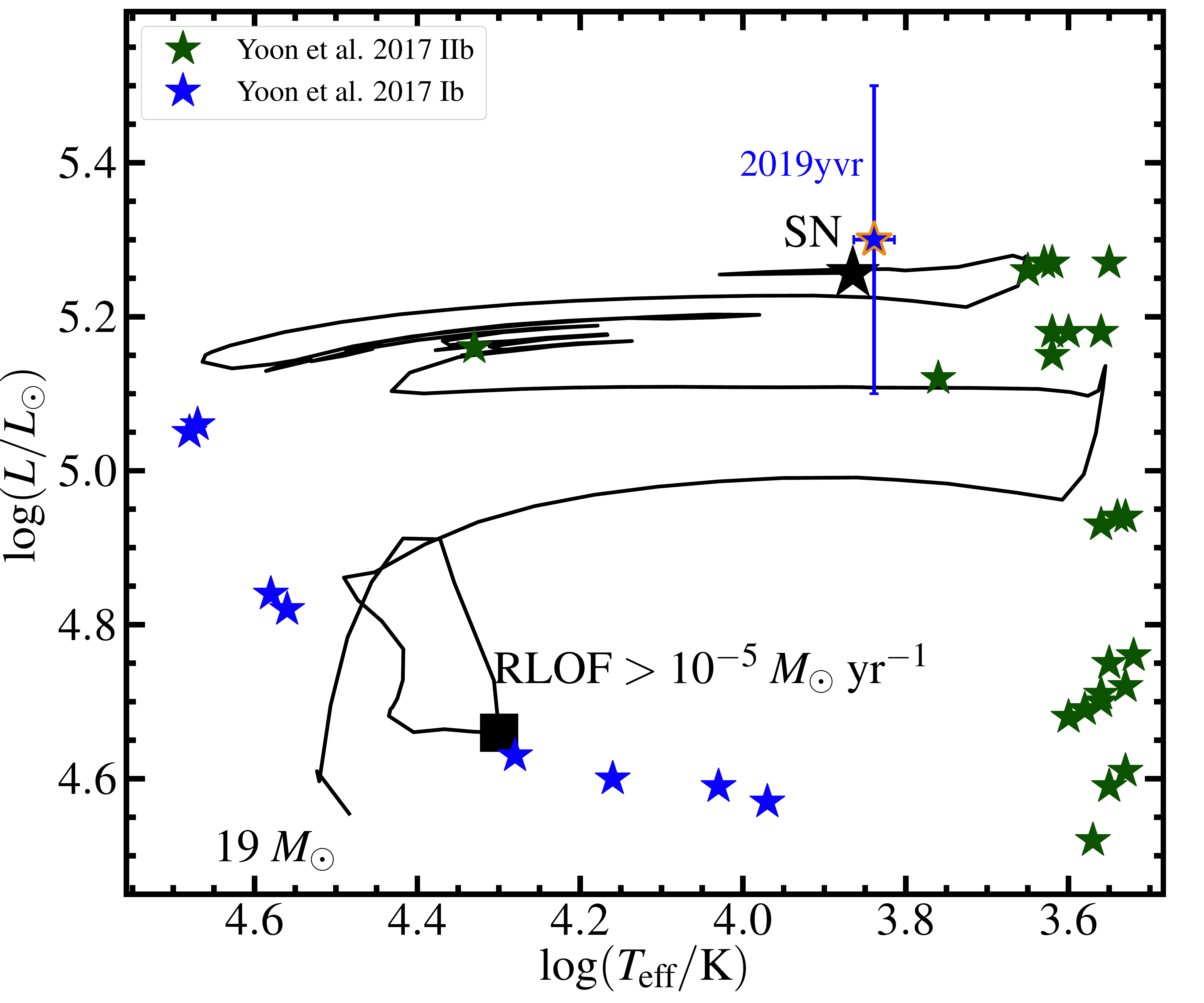}
    \caption{({\it Top}): A Hertzpsrung-Russell diagram showing the location of the SN~2019yvr progenitor candidate (blue star) with comparison to SN~IIb progenitor stars (green squares), iPTF13bvn \citep[blue diamond;][]{Cao13,Bersten+14,Folatelli16}, and SN~II progenitor stars \citep[red squares;][]{Smartt15}.  We overplot MIST single-star evolutionary tracks from \citet{choi+16,choi+17} for comparison.  {\it Bottom}: A 19+1.9~$M_{\odot}$ binary star evolution track from BPASS v2.2 \citep{eldridge+17}, which is consistent with the pre-explosion photometry of the SN~2019yvr progenitor candidate.  We highlight the location on the track where the primary star begins Roche-lobe overflow (RLOF; square), reaches its minimum hydrogen envelope mass ($0.047~M_{\odot}$, circle), and terminates as a supernova (star).  We also show binary star models from \citet{yoon+17} with outcomes predicted for type Ib (blue) and IIb SNe (green).}
    \label{fig:hr}
\end{figure}

Given our analysis in \autoref{sec:candidate}, we assume throughout this discussion that the SN~2019yvr pre-explosion counterpart is dominated by emission from the SN progenitor system.  From our inferences about this source above as well as our knowledge of SN~2019yvr, we consider what evolutionary pathways could lead to the source observed in the {\it HST} photometry as well as the resulting SN.  These pathways need to explain several facts referenced throughout the previous analysis, which we summarize here as:

\begin{enumerate}[left=0pt]
    \item SN~2019yvr was a SN~Ib with no evidence for hydrogen in its early-time spectra, starting from 7~days before peak light \citep{ATel13375} until well after peak light.  Following models in \citet{dessart+12}, this suggests that the progenitor star must have had $<$10$^{-3}~M_{\odot}$ \citep[but possibly as much as $0.03~M_{\odot}$;][]{Hachinger12} of hydrogen remaining in its envelope at the time of explosion.
    \item SN~2019yvr began interacting with CSM starting around $150$~days after explosion and exhibited strong H$\alpha$, radio, and X-ray emission consistent with a shock formed in hydrogen-rich material \Auchp.  Using conservative assumptions about the SN shock velocity (10,000~km~s$^{-1}$) and velocity of the CSM (100~km~s$^{-1}$), we infer that this material must have been ejected at least $\approx$44~yr prior to core collapse and implying a shell or clump of material at $>$1000~AU from the progenitor system (see estimate from \autoref{eqn:mle} in \autoref{sec:eruptions}).  Although we find it more likely that this material came from the SN progenitor star itself given the frequency of SNe~Ib with late-time interactions \citep[i.e., similar to SN~2014C;][]{Milisavljevic15,Margutti+16}, it is possible that this material came from a binary companion and the actual SN~2019yvr progenitor star was hydrogen free on this timescale.  Beyond this interaction, there is no evidence for dense CSM in early-time spectra or light curves, all of which appear similar to SNe~Ib without evidence for these interactions.
    \item SN~2019yvr exhibited a large line-of-sight extinction ($A_{V}\approx2.4$~mag) while its Milky Way reddening was only $E(B-V)=0.02$~mag.  We infer from the strong Na\I~D line at the redshift of the SN~2019yvr host galaxy NGC~4666 that this extinction implies a significant dust column in the NGC~4666 interstellar medium toward SN~2019yvr and/or a correspondingly large column of circumstellar dust in the environment of the progenitor system itself.  There is no clear evidence for any mass ejections or warm circumstellar gas in a compact shell around the progenitor system, either in pre-explosion data or from circumstellar interaction once SN~2019yvr exploded.
    \item There is a single, point-like progenitor candidate to SN~2019yvr detected in pre-explosion \hst\ imaging.  This source exhibits very little photometric variability over a 110~day period from 2.7 to 2.4~yr prior to core collapse.  Before applying our host extinction estimate but applying Milky Way extinction, this progenitor candidate has a red colour of $m_{\mathrm{F555W}}-m_{\mathrm{F814W}}=1.065$~mag.  Accounting for the inferred extinction and distance modulus above, the source is relatively luminous with $M_{\mathrm{F555W}}=-7.8$~mag (roughly in $V$-band).  This value is consistent with massive stars but low for a stellar cluster \citep[as in][]{Gieles06}.
    \item Accounting for all extinction, the progenitor candidate is consistent with a \Ldot=5.3$\pm$0.2 and $T_{\mathrm{eff}}\approx6800$~K star, which implies a photosphere with a radius of $\approx320~R_{\odot}$ at 2.6~yr prior to the SN~2019yvr explosion.  Such a star would be closest in temperature and luminosity to yellow supergiants confirmed as SN~IIb progenitor stars (\autoref{fig:hr} and green circles for SNe~1993J, 2008ax, 2011dh, 2013df, and 2016gkg).  Comparing to stellar SEDs, the spectral type and luminosity class are best matched to F4--F0 supergiant stars.
    \item Comparing the pre-explosion photometry to BPASS binary stellar evolution tracks in \citet{eldridge+17}, the best-fitting model is a 19~$M_{\odot}$+1.9~$M_{\odot}$ system that undergoes common envelope evolution and strips most of the material from the progenitor star.  Immediately prior to explosion, the primary star retains 0.047~$M_{\odot}$ of hydrogen in its envelope, inconsistent with the masses for SN~Ib systems given above.  No other BPASS models were consistent with both our pre-explosion photometry and a system that produced a SN explosion.
\end{enumerate}

\subsection{The Anomalously Cool Progenitor and the Type Ib Classification of SN~2019yvr}\label{sec:cool}

We compare the SN~2019yvr progenitor candidate in a Hertzsprung-Russell diagram to other known progenitor systems in \autoref{fig:hr}, including SNe~II \citep[][and references therein]{Smartt15}, SNe~IIb \citep[SNe~1993J, 2008ax, 2011dh, 2013df, and 2016gkg;][]{aldering+94,crockett+08,maund+11,vandyk+14,kilpatrick17:16gkg}, and the progenitor of the SN~Ib iPTF13bvn \citep{Cao13,Bersten+14,Folatelli16}.  We also note that there is a single SN~Ic progenitor candidate for SN~2017ein \citep{Kilpatrick18:17ein,vandyk+18}, but the source has a luminosity \Ldot$\approx$6.0 and temperature $T_{\mathrm{eff}}>10^{5}$~K and so is off our plotting range (and may actually be a very young open cluster).  For comparison, we also overplot single-star evolutionary tracks from the Mesa Isochrones \& Stellar Tracks code \citep{choi+16,choi+17}.

The most notable feature of the SN~2019yvr progenitor candidate in \autoref{fig:hr} and compared with other stripped-envelope SN progenitor stars is its relatively cool effective temperature, which in turn implies an extended photosphere given our constraints on its SED.  Assuming a single-star origin and judging solely by the source's inferred luminosity and temperature of \Ldot=5.3 and $T_{\mathrm{eff}}=6800$~K, the counterpart is consistent with a $30~M_{\odot}$ star in the so-called ``Hertzsprung gap'' \citep[see, e.g.,][]{deJager97,Stothers01}.  This is in sharp contrast to the predicted progenitors of SNe~Ib, which are thought to have low-mass, compact envelopes, consistent with a star that has almost no hydrogen in its outer layers \citep[][]{yoon+15}.

Given these facts, we consider the implications of different progenitor scenarios.  In particular, we emphasize the apparent contradiction between a SN from a star without a significant hydrogen envelope mass and a pre-explosion counterpart consistent with a massive star with a significantly extended photosphere, which typically requires a non-negligible hydrogen envelope mass.  Combined with the SN~2014C-like circumstellar interaction observed at late times, SN~2019yvr may offer significant insight into mass loss in late-stage stellar evolution for stripped-envelope SNe.  Here we review ``standard'' single and binary star scenarios and assess whether they can explain all of these properties.

\subsubsection{Single-star Models}

It is still debated if single massive stars can evolve to the point where they would explode as yellow supergiants, in part because the Hertzsprung gap is typically a short-lived and transitional phase in the post-main sequence.  Standard single-star evolution would suggest that a star with an effective temperature of $T_{\mathrm{eff}}=6800$~K and \Ldot=5.3 retains a massive hydrogen envelope: a $30~M_{\odot}$ initial mass star would retain a $60\%$ surface hydrogen mass fraction on its first passage through the Hertzsprung gap \citep[following the structure of model stars in][]{choi+16}.  Thus, in the context of stripped-envelope SNe, any single-star model would likely enter the yellow supergiant phase after evolving through the RSG branch and shedding the remainder of its hydrogen envelope. \citet{Georgy09} demonstrate that such evolution is possible if RSG mass loss rates are increased by approximately an order of magnitude towards the end of their nuclear lives.

While the physics of an enhanced late-stage mass loss is unclear (see, e.g. \citealt{Yoon2010}), there do exist a number of luminous yellow supergiants (termed ``yellow hypergiants'') which are hypothesized to be such post-RSG stars. Many yellow hypergiants are extremely variable with high mass-loss rates, such as $\rho$~Cas \citep{Smith14,Lobel15} and V509~Cas \citep{Percy92}, and are located toward the end of the luminous blue variable (LBV) bistability track \citep{smith+04}. This variability involves a rapid (years to decades) evolution between quiescent, hot phases and erupting, cool phases with extreme mass-loss episodes \citep[e.g., $10^{-4}~M_{\odot}~\mathrm{yr}^{-1}$ as in the yellow hypergiants $\rho$~Cas and HR~8752;][]{deJager98,Humphreys02}.

Yellow hypergiants undergoing extreme mass loss have been proposed as candidates for type Ib/c progenitors as their LBV-like mass ejections provide an efficient way to rid the star of its hydrogen and helium envelope in the years before core collapse \citep[as in SN~2006jc;][]{Foley07,smith+08b}. The presence of these stars in the gap and the fact that some of them explode as type IIb SNe \citep[the bluest SN~IIb progenitor stars span this gap, e.g.,][]{crockett+08,kilpatrick17:16gkg} strengthens the association between extreme mass loss and stripped-envelope SNe \citep{deJager98,Stothers99}. While it is debated if such enhanced mass-loss rates are possible for single stars on the RSG branch \citep[e.g.,][]{Beasor20}, mass-loss episodes and variability of stars near this point on the Hertzpsrung-Russell diagram suggests it is possible to rapidly shed their hydrogen envelopes and increase in temperature over timescales of years \citep[as observed with HR~8752, for example;][]{deJager97}. However, yellow hypergiants that we observe in this region of the Hertzsprung-Russell diagram have massive, hydrogen-rich envelopes and their winds are known to be hydrogen rich \citep{smith+04}.  Assuming such a star exploded as a SN~Ib only 2.6~yr after being observed in this evolutionary phase, it would either need to shed its remaining envelope in the final 2.6~yr or remain cool after retaining only a trace hydrogen envelope.  The former scenario will be discussed in \autoref{sec:eruptions}, below.

Alternatively, it is worth considering if a single star with virtually no remaining hydrogen in its envelope could inflate to a radius of 320~$R_{\odot}$ while exhibiting a photospheric temperature of 6800~K---as required for SN~2019yvr and its progenitor candidate.  The coolest known helium stars are only $\approx$10,000~K and these examples are significantly less luminous than the observed counterpart.  For example, LSS~4300 is $T_{\mathrm{eff}}= 11,000$~K \citep{Drilling84,Schon84}, KS Per is 10,000~K \citep{Woolf73,Drilling82}, and $\nu$~Sgr is 11,800~K \citep{Frame95}.  The counterpart we observe is only this hot assuming our SED modeling is inconsistent with the true temperature at $>3\sigma$ or if there is $>$2~mag more extinction than we assumed at 2.6~yr before explosion.  Even if we assume either of these scenarios is true, the progenitor candidate would still have \Ldot$\geq$5.3, which is much more luminous than the most luminous helium stars that reach \Ldot=4.6 \citep[see previous references and][]{Dudley92}. Thus an anomalously high-mass and luminous helium star would be needed to match to SN~2019yvr, which in general is not allowed by standard single star stellar evolution models.

\subsubsection{Binary Star Models}\label{sec:disc-binary}

As shown above, the best-fitting binary systems for our SN~2019yvr pre-explosion photometry are short-period, low mass ratio systems that undergo CEE.  While such systems are expected to have complex circumstellar environments, possibly consistent with SN~2019yvr, we emphasize that all of the BPASS models terminate with final hydrogen-envelope masses of $>$0.038~$M_{\odot}$.  This is inconsistent with a classification as a SN~Ib via the models of \citet{dessart+12} and \citet{Hachinger12}.  In addition, there is debate as to whether the evolution of such high mass ratio binaries within BPASS is representative of actual stellar systems.
\citet{Neugent2018,Neugent2020} argue that any star that will eventually explode as a core-collapse SN evolves off the main sequence fast enough that companion stars with initial masses $\lesssim$3~$M_{\odot}$ would not have enough time to complete their contraction phase and thus would still be protostars. This is not accounted for in current BPASS models, which initialize all stellar masses on the ZAMS simultaneously.

More broadly, the progenitor systems of SNe~Ib are predicted to be low-mass helium stars that evolve via binary evolution. While they can expand to moderately large radii due to shell burning, they typically remain $<$100~$R_{\odot}$ \citep[][]{kleiser+18,Laplace20}.  Their photospheric temperatures are thus significantly hotter than the \hst\ photometry of the SN~2019yvr progenitor candidate implies. The latter allows at most 11,000~K as opposed to the $\approx$45,000~K photosphere of the progenitor star to iPTF13bvn \citep{eldridge+16,Folatelli16}.  \citet{yoon+15} predict that the best observational counterparts to SN~Ib progenitor systems are binary, stripped WR stars similar to HD~45166 \citep{groh+08} and WR7a \citep{Oliveira+03}, both of which have $T_{\mathrm{eff}}>20$,000~K.  Nearly all of these stars are known to be in binaries with close ($<$few day) orbits where the primary has been stripped to $3$--$10~M_{\odot}$, likely through Case B mass transfer \citep{yoon+12,yoon+15,yoon+17}.  The companion can span a wide range of luminosities and evolutionary states \citep[e.g., WR7a exhibits a 0.204~day binary orbital period, but no secondary is observed implying a very low-mass star or a compact object;][]{pereira+98,Oliveira+03}.  The primary stars are left completely stripped of a hydrogen envelope. This is in contrast to the best-fitting BPASS model, which undergoes extreme mass loss due to CEE and RLOF during the post-main sequence but stops before the hydrogen envelope is completely depleted \citep{Delgado81,Ivanova13,eldridge+17}.

Thus, as with the single star models, there remains significant tension as to whether any standard binary evolution model can reproduce both the progenitor candidate and a system that explodes as a SN~Ib.  We are left needing to invoke some additional mechanism that can account for both the cool photospheric temperature 2.6~yr prior to explosion and the negligible hydrogen envelope mass at the time of core collapse. In the following sections we discuss two potential resolutions to this paradox (\autoref{sec:eruptions} and \autoref{sec:rtau}) and highlight some evolutionary scenarios that may be allowed while also explaining the dense CSM shell observed around SN\,2019yvr (\autoref{sec:scenarios}).

\subsection{Ejection of the Final Hydrogen Envelope in the Last 2.6 Years Prior to Core Collapse}\label{sec:eruptions}

One way to resolve the apparent conflict between the extended progenitor radius and lack of hydrogen in the ejecta of SN\,2019yvr would be if the progenitor star \emph{did} possess a envelope with $\sim$0.01--0.03~M$_\odot$ of hydrogen (similar to the yellow supergiant progenitor stars of SNe~IIb) at the time of the \hst\ observations, but somehow managed to lose this material in the intervening 2.6~yr. Such a scenario is not unprecedented: episodic mass ejections have been invoked to explain stripped-envelope SN progenitor systems that exhibit dense shells or clumps of CSM \citep[see, e.g.,][]{Chugai06,Bietenholz14,Chandra17,Mauerhan18,Pooley19,Sollerman20,Tartaglia20}.  This was also observed directly for SN~2006jc, a SN~Ib with a pre-explosion outburst 2~yr before explosion \citep{Foley07,smith+08b,pastorello+08,Maund16}, which later manifested as circumstellar interaction.  Indeed, SN\,2019yvr shows evidence for relatively narrow H$\alpha$, radio, and X-ray emission $\approx$150~days after core collapse, providing evidence for episodic mass ejections throughout the progenitor's final evolutionary stages.

In this scenario, this hydrogen-rich material should be located at some radius around the progenitor star. A key question for SN\,2019yvr is whether the shell of H-rich material encountered by the SN ejecta $\sim$150 days post-explosion could be the remnants of such an ejection that occurred between the {\it HST} observations and explosion.
While the location and timing of this ejection will be discussed in detail in \Aucht, we perform an order of magnitude calculation to investigate this possibility here.  For a CSM wind velocity ($v_w$) and a SN shock velocity ($v_s$) the detection of CSM interaction starting $\sim$150~days after core collapse implies a mass-loss event that occurred:

\begin{equation}
t_{\rm mle}\approx 44 \left( \frac{v_s}{ 10^{4}{\rm ~km s^{-1}}}\right) \left(\frac{ v_{\mathrm{w}}}{100{\rm ~km s^{-1}}}\right)^{-1}\;{\rm yrs}\label{eqn:mle}
\end{equation}

\noindent before core collapse. We have scaled our results to an average shock velocity of 10,000~km~s$^{-1}$ and wind velocity of 100~km~s$^{-1}$. The former is roughly consistent with the velocity inferred from helium absorption in our spectra of SN~2019yvr near maximum light, which is a lower limit for the shock velocity. While the latter is less constrained, an ejection speed of $\sim$100 km s$^{-1}$ is approximately the escape speed for a star with a radius of 320~$R_{\odot}$ (as inferred from our progenitor candidate) and a mass of 5$-$10 M$_{\odot}$. For these assumptions, the mass loss event would have occurred significantly earlier than the time of the {\it HST} observations.

Therefore, if the hydrogen we observe in CSM is the same material inferred from the progenitor star photosphere, the wind velocity must be at least 1800~km~s$^{-1}$ such that the progenitor star could eject it after the \hst\ data were obtained.  A wind at this speed can only be achieved by a compact and massive WR-like star \citep[similar to those presented in][]{Rochowicz95,vanderhucht+01}, inconsistent with our pre-explosion photometry. Such a high velocity would therefore require and ejection mechanism capable of accelerating material to $\gtrsim$10$\times$ the stellar escape speed. This would be in contrast to current theoretical models for both wave-driven mass loss in hydrogen-poor stars (which have terminal velocities of a few hundred km~s$^{-1}$ \citealt{fuller+18}) and common-envelope ejections (which tend to proceed at roughly the escape velocity).

Unless a stronger ejection mechanism can be identified, while the mass-loss event that led to the material at $\approx$1000~AU was timed soon before the explosion of SN~2019yvr, it may not be directly associated with the relatively cool photosphere we infer from the \hst\ imaging.  Multiple eruptive mass-loss events would be needed to explain both the CSM and the final depletion of the progenitor star's hydrogen envelope assuming the pre-explosion counterpart, the source of the material around SN~2019yvr, and the SN~2019yvr progenitor star are all the same.  The second ejection would place material closer to the progenitor star, which is not detected in our light curves or spectra \citep[e.g., via enhanced emission due to CSM interaction or flash ionisation features similar to][]{gal-yam+14}.  It is possible that this material was missed due to a delay between the explosion of SN~2019yvr and its discovery, and detailed analysis of these data will be carried out in \Aucht to assess whether or not this could be the case.

\subsection{A Quasi-photosphere Seen as an Inflated Star}\label{sec:rtau}

If the scenario discussed in \autoref{sec:eruptions} is not viable, then the SN\,2019yvr progenitor candidate must have contained virtually no hydrogen 2.6~yr before explosion \citep[$<10^{-2}~M_{\odot}$ as required for a type Ib classification by][]{dessart+12,Hachinger12}. In this case, we would require that the envelope was inflated by some process not accounted for in standard models of stellar evolution. Here, we consider two scenarios in which a star with only a trace hydrogen envelope could exhibit a photospheric radius of $\approx$320~$R_{\odot}$ or roughly 1.5~AU: (a) formation of a pseudo-photosphere in a dense stellar wind, and (b) inflation due to an additional heat/energy source that produces a radiation pressure supported envelope.

\subsubsection{A Stellar Wind}

For certain mass-loss rates, a stellar wind can become optically thick and appear at a radius well beyond that of the underlying star \citep[see, e.g.,][]{Gallagher92}.  The radius at which this quasi-photosphere forms ($R_{\tau}$) depends on the density ($\rho$) and opacity ($\kappa$) in the wind, with optical depth ($\tau$) expressed as

\begin{equation}
    \tau = \int_{R_{\tau}}^{\infty} \rho \kappa dr.
    \label{eq:tau}
\end{equation}

\noindent From this equation, a quasi-photosphere will form when $\tau \gtrsim 1$.  Assuming that the radial density profile in the wind follows $r^{-2}$ for a constant mass-loss rate ($\dot{M}$), these conditions are largely dependent on the properties of the wind and thus the underlying star.  From equation (16) in \citet{deKoter96}, where the wind opacity is modeled as temperature-dependent bound-free opacity from the Paschen continuum, the quasi-photosphere constraint above implies

\begin{multline}
    \dot{M} > 1.9\times10^{-4}~M_{\odot}~\mathrm{yr}^{-1}
     \left(\frac{T_{\rm eff}}{10^{4}~\mathrm{K}} \right)^{3/4}
     \left(\frac{R_{\tau}}{300~R_{\odot}}\right)^{3/2} \\
     \left(\frac{v_{w}}{100~\mathrm{km~s}^{-1}}\right)^{1/2}
     \left(\frac{c_{s}}{10~\mathrm{km~s}^{-1}}\right)^{1/2}
   \label{eq:cond1}
\end{multline}

\noindent with $v_{w}$ the wind velocity and $c_{s}$ the local sound speed.

To estimate mass loss rate required to explain the photosphere observed for the progenitor candidate of SN\,2019yvr, we assume that the wind could be as fast as $100$~km~s$^{-1}$,as described in \S~\ref{sec:eruptions} (a detailed analysis of the CSM properties will be presented in \Aucht).  We also use our constraint on the effective temperature of the observed photosphere $T_{\mathrm{eff}}=6800$~K and radius $R=320~R_{\odot}$ above, which gives a local sound speed $c_{s}\approx7$~km~s$^{-1}$ and implies $\dot{M}>1.3\times10^{-4}~M_{\odot}~\mathrm{yr}^{-1}$.

This mass-loss rate is extreme for a star with $\Ldot = 5.3$ \citep[even yellow hypergiants or low-luminosity LBVs as in][]{smith+04}, which tend to have time-averaged $\dot{M}\approx10^{-5}~M_{\odot}~\mathrm{yr}^{-1}$ \citep{humphreys+94}.  However, the total mass of material needed to form a quasi-photosphere is only $\dot{M} v_{w}^{-1} R_{\tau} \approx 10^{-5}~M_{\odot}$.  This material can easily be shed from the star's remaining hydrogen/helium envelope over a brief period of strong mass loss, although the timescale for such a shell to form would be only $R_{\tau}/v_{w}=26$~days.

However, we do not observe any emission lines or ``flash spectroscopy'' features in our early spectra of SN\,2019yvr as may be expected if dense wind material is present.  This scenario would also require a significant change in the mass loss from the SN~2019yvr progenitor timed $\lesssim$3 years prior to explosion, which is closely aligned with predictions for instabilities during oxygen burning \citep{meakin+06,Arnett14}.  However, there is no other evidence for a dense wind that could form the observed photosphere.

\subsubsection{A Radiation Dominated Envelope}

An alternative is that the progenitor star retained a small amount of hydrogen \citep[$<10^{-2}~M_{\odot}$ as required for a type Ib classification by][]{dessart+12,Hachinger12} with an envelope that has relaxed to a steady-state configuration following some injection of additional energy. Assuming that radiation pressure  dominates over gas pressure, the radial density profile in the wind will follow $r^{-3}$ \citep{ulmer1997}. Hydrostatic equilibrium holds up to the radius at which the radiation is no longer trapped, and so we associate the outer radius of this envelope with the photospheric condition $\tau=1$ as

\begin{equation}
R_{\tau}\approx 300 \left(\frac{M_{\rm env}}{ 10^{-4}M_{\odot}}\right)^{1/2} R_{\odot},
\end{equation}

\noindent where we assume Thompson opacity and a hydrogen envelope with a Solar mass fraction. We note that an inflated progenitor could exist provided that the mass contained in the envelope is $>10^{-4}M_{\sun}$, which is both reconcilable with a type Ib classification for SN\,2019yvr and in general is less restrictive than the condition given by \autoref{eq:cond1}.

However, even in this case several mysteries remain as to the nature of the progenitor system.  In particular, for a complete description of SN\,2019yvr, the progenitor system would still need to eject a shell of hydrogen-rich material at $\sim$1000 AU. This will be be discussed in \autoref{sec:scenarios}.

\subsection{Progenitor Scenarios for SN~2019yvr}\label{sec:scenarios}

In the sections above, we have argued that reconciling the extended progenitor radius and lack of hydrogen in the ejecta of SN2019yvr requires a system that either:

\begin{enumerate}[left=0pt]
    \item ejected the remainder of its hydrogen envelope in the final 2.6 yrs pre-explosion (\autoref{sec:eruptions}).
    \item undergoes a process not accounted for in standard models of evolution and that leads to additional inflation or has sufficient material around the progenitor to form a quasi-photosphere in the CSM (\autoref{sec:rtau}).
\end{enumerate}

In addition, in either case, a viable progenitor scenario for SN\,2019yvr must also explain a shell of hydrogen-rich material observed at $\sim$1000 AU that was likely ejected 50--100 years prior to explosion. Below we consider two pathways that may be consistent within these constraints. We emphasize that this is not an exclusive list of possible scenarios. No existing model currently explains all the observed properties of SN\,2019yvr and its progenitor system.

\subsubsection{A Luminous Blue Variable Phase}\label{sec:lbv}

One evolutionary pathway that naturally explains many of the observables for SN~2019yvr is a series of eruptions from a massive star in a LBV phase.  These eruptions are known to precede many SNe, some of which are thought to be core-collapse explosions \citep[as was argued for SN~2009ip][]{mauerhan+13}, although this interpretation of LBVs as a phase immediately preceding core collapse remains controversial \citep[][]{margutti+14}.

Many types of progenitor systems exhibit pre-SN eruptions in the final years to weeks before explosion, notably for the progenitors of SNe~IIn \citep[SNe with narrow Balmer lines in their optical spectra, indicative of interaction between ejecta and a dense mass of CSM;][]{Smith17}.  These systems must eject from 0.1--10~$M_{\odot}$ over short timescales \citep{smith+07,fox+10,fox+11} requiring an extreme mass-loss rate and variability on the timescale of the eruptions.  Their progenitor systems have also been observed in the literature, notably for SN~2009ip whose progenitor star had an initial mass $>$60~$M_{\odot}$ \citep{smith+09,foley+11}.  Although there is some ambiguity whether these events are actually terminal explosions of massive stars \citep{mauerhan+13,margutti+14}, the eruptive mechanism that fills their environments with dense shells of CSM may be more common among stars at a wide range of initial masses.  Indeed, so-called SN impostors \citep[lower luminosity, likely non-terminal explosions of massive stars that resemble SNe~IIn;][]{VanDyk00,maund+06,pastorello+10,ofek+16} come from systems with initial masses possibly as low as 20~$M_{\odot}$ \citep[][]{Kilpatrick18:16cfr}.  Such an eruption for SN~2019yvr and SN~2014C-like events \citep[hypothesized by][]{Milisavljevic15} could explain the source of the CSM, although most SN~IIn and SN impostor progenitor stars retain some hydrogen leading to broad Balmer lines in their spectra.

For SN~2019yvr, we would require a scenario analogous to SNe~IIn and Ibn with strong circumstellar interaction soon after explosion \citep{Smith17,Hosseinzadeh17} but involving multiple mass-loss episodes timed years or even decades ahead of core collapse instead of months to years.  The main caveat in invoking this progenitor scenario is whether episodic mass ejections can fully strip the progenitor star's hydrogen envelope on the timescale required by our \hst\ observations.  While LBV eruptions provide a compelling mass-loss scenario as progenitor systems extend nearly to the parameter space of the Hertzsprung-Russell diagram where we find the SN~2019yvr counterpart \citep[see SN~2009ip, 2015bh, Gaia16cfr;][]{smith+09,mauerhan+13,elias-rosa+16,thone+17,Kilpatrick18:16cfr}, these events tend to result in hydrogen-rich explosions with long-lived Balmer emission.  The final eruptions of SN~2019yvr would need to strip a sufficiently small amount of hydrogen (potentially hundredths of a Solar mass as in \autoref{sec:binary}) that interaction with this material would not be observed in early time spectra.  Detailed analysis of the initial observations of SN~2019yvr could place stronger constraints on whether this scenario occurred.

\subsubsection{A Common Envelope Mass Loss Scenario}\label{sec:ce}

An alternative evolutionary pathway in the final decades of evolution for the SN~2019yvr, and put forward for SN~2014C in \citet{Margutti+16}, is CEE leading to ejection of the primary star's hydrogen envelope $<$100~yr before explosion.  In general, this is not expected as RLOF and CEE are commonly associated with mass transfer much earlier in the primary star's life cycle, for example, immediately after helium core contraction as discussed in \autoref{sec:introduction}.  The progenitor would need to be inflated in a later evolutionary stage after helium burning to restart mass transfer \citep[i.e., through Case C mass transfer, see][]{Schneider15}.  This is predicted to occur for only $\sim$5$-$6\% of binary system with primary masses between 8$-$20 M$\odot$ \citep{Podsiadlowski1992}.

However, if the primary star's envelope can inflate during this phase, the companion will spiral inward and start CEE, resulting in the ejection of a significant fraction of the envelope over $<1$~yr \citep{Ivanova13}.  The velocity of that material would be comparable to the escape velocity of the primary, and so would likely be $<100$~km~s$^{-1}$ depending on the envelope structure of the primary star at onset of CEE.  In this way, the material could survive long enough that the SN ejecta can encounter it within the first year after core collapse, as observed at $\sim$150 days in SN\,2019yvr.

The key question for SN~2019yvr is how the post-CEE system evolves in the final years before core collapse such that it resembles the pre-explosion counterpart and also explodes as a SN~Ib.  Case-C CEE is an appealing solution to this problem as the remaining envelope is predicted to become unstable, leading to dynamical pulsations and steady mass loss in the remaining years before core collapse \citep[especially for extremely luminous stars;][]{Heger97}.  Thus a post-CEE star could still form a quasi-photosphere assuming the mass-loss rate exceeded 10$^{-4}~M_{\odot}~\mathrm{yr}^{-1}$ as in \autoref{sec:rtau}. Alternatively, the inspiral during CE phase itself supplies a source of heat in the stellar envelope. Thus, after the ejection of most of its hydrogen envelope, the resulting post-CEE hydrogen-deficient envelope could remain partly inflated.  If the resulting star was dynamically unstable and losing mass significantly faster than the $\approx10^{-5}~M_{\odot}~\text{yr}^{-1}$ predicted for yellow hypergiants \citep{humphreys+94}, it could rapidly shed even the $10^{-2}$--$10^{-1}~M_{\odot}$ that is predicted to remain in the envelope of comparable stars from BPASS.  This would simultaneously explain the circumstellar material as the result of CEE, the apparently cool photosphere, and the lack of hydrogen in the primary star's envelope at core collapse.

\subsection{Comparisons to Progenitor Star Constraints for SN~2014C and SNe~Ib with Late-time Circumstellar Interaction}

SNe~Ib with late-time circumstellar interaction similar to SN~2019yvr are not unprecedented and may in fact represent a large fraction of stripped-envelope SNe overall.  As discussed in \autoref{sec:introduction}, SN~2014C began interacting with its circumstellar environment a few weeks after explosion \citep{Margutti+16}, and both SN~2019yvr and SN~2014C have relatively deep progenitor constraints considering both have pre-explosion \hst\ imaging and occurred at $\approx$15~Mpc \citep[][]{Milisavljevic15}.  More broadly, there are numerous examples of stripped-envelope SNe with circumstellar interaction soon after explosion \citep{Chugai06,Bietenholz14,Chandra17,Mauerhan18,Pooley19,Sollerman20,Tartaglia20}, and so any mechanism that we invoke above may need to explain the presence of CSM for a significant fraction of all core-collapse SNe \citep{Margutti+16}.

A compact star cluster toward the position of SN~2014C in pre-explosion imaging \citep{Milisavljevic15} had a best-fitting age in the range $30$--$300$~Myr, although it could be as young as $10$~Myr.  This age implies a main-sequence turnoff mass of $3.5$--$9.5~M_{\odot}$ depending on the metallicity of the cluster.  Assuming this cluster hosted the SN~2014C progenitor star and considering the fact that stars in this mass range either do not explode as core-collapse SNe or are thought to lead to SNe~II, \citet{Milisavljevic15} inferred that a more likely explosion scenario for SN~2014C was through binary star channels for a star with a LBV-like phase and $M_{\mathrm{ZAMS}}>$20~$M_{\odot}$ (implying that the cluster is younger than its colours suggest).  Given the photometric and spectroscopic similarity between SNe~2014C and 2019yvr and the presence of a strong density gradient in the circumstellar environments of both, it is possible that both systems could be explained through eruptive, LBV-like mass loss or CEE as discussed above.

However, these evolutionary phases would require that the SN~2019yvr progenitor star had extreme photometric variability, and the lack of any such variability in the pre-explosion photometry indicates that these episodes would have occurred outside the short window in which we constrain the progenitor star's evolution.  Assuming that the mass-loss episode was driven by an instability in late-stage nuclear burning \citep[e.g.,][]{Arnett14}, the star could also be inflated temporarily.  \citet{Smith14b} suggest this inflation would occur on a timescale comparable to the orbital period in binary systems, but we see no signature of this variability on $30$--$100$~day timescales.  Deep, high-cadence limits, such as those from the Young Supernova Experiment \citep{Jones19,Jones20} or the upcoming Vera C. Rubin Observatory Legacy Survey of Space and Time \citep[down to r$\approx$24.5~mag on a 3--4 day cadence;][]{Ivezic19}, will be able to detect or rule out these mass-loss episodes for future events, potentially constraining the progenitor scenarios for stripped-envelope SNe in the volume where such variability is detectable.

\section{Conclusions}\label{sec:conclusions}

We present pre-explosion imaging, photometry, and spectroscopy of the SN~Ib 2019yvr.  We find:

\begin{enumerate}[left=0pt]
    \item SN~2019yvr was a SN~Ib with a large line-of-sight extinction spectroscopically similar to iPTF13bvn.  SN~2019yvr exhibited signatures of interaction $150$~days after discovery consistent with a shock between SN ejecta and dense, hydrogen-rich CSM \Auchp\ and similar to SN~2014C \citep{Milisavljevic15,Margutti+16}.  This interaction suggests that the SN~2019yvr progenitor star underwent an eruptive mass-loss episode at least 44~years before explosion.
    \item There is a single source in {\it Hubble Space Telescope} imaging obtained $\approx$2.6~year before discovery and consistent with being the progenitor system of SN~Ib~2019yvr.  Comparing to blackbodies, single-star SEDs, and binary star models, we find that the SN~2019yvr progenitor star is consistent with a star with $\Ldot = 5.3 \pm 0.2$, $T_{\mathrm{eff}} = 6800\substack{+400\\-200}$~K, and thus close ($<$5~day period) binary-star models with initial masses around $19~M_{\odot}$.  However, the cool effective temperature and high luminosity implies a remaining hydrogen envelope mass of at least $0.047~M_{\odot}$ in the binary star model, which is inconsistent with the lack of hydrogen in spectra of SN~2019yvr.
    \item Comparison to SN~Ib progenitor candidates indicates that SN~2019yvr is much cooler than what is predicted for a hydrogen-stripped star, much more similar to the identified progenitors of type IIb SNe.  Overall, the progenitor candidate appears cool and inflated relative to the progenitor of iPTF13bvn and helium stars.
    \item We infer that an extreme and episodic mass-loss scenario is required to produce both a stripped-envelope SN progenitor system and the luminous, cool progenitor candidate.  The binary evolution scenarios discussed above do not incorporate physical scenarios that can lead to extreme or eruptive mass loss soon before explosion, and barring such a mass-loss scenario they do not produce a star whose hydrogen envelope is consistent with the SN~Ib classification.  We propose that LBV-like mass ejections or CEE provide natural explanations for the stellar classification, the lack of a massive hydrogen envelope, and the presence of dense CSM.  We hypothesize that if this mass-loss mechanism occurs, the star could have formed a quasi-photosphere from CSM in its environment, requiring either a mass loss at a rate $>1.3~M_{\odot}~\mathrm{yr}^{-1}$ or a radiation supported hydrogen envelope with a mass $>10^{-4}M_\odot$ at 2.6~yr before core collapse.
\end{enumerate}

\bigskip\bigskip\bigskip
\noindent {\bf ACKNOWLEDGMENTS}
\smallskip
\footnotesize

We thank J.J. Eldridge and H. Stevance for helpful comments about our BPASS analysis, J. A. Vilchez, A. Campillay, Y. K. Riveros and N. Ulloa for help with the Swope observations, as well as R. Carrasco for support of our Gemini-S/GSAOI programme.
C.D.K.\ acknowledges support through NASA grants in support of {\it Hubble Space Telescope} programmes GO-15691 and AR-16136.
M.R.D.\ acknowledges support from the NSERC through grant RGPIN-2019-06186, the Canada Research Chairs Program, the Canadian Institute for Advanced Research (CIFAR), and the Dunlap Institute at the University of Toronto.
K.A.\ is supported by the Danish National Research Foundation (DNRF132) and VILLUM FONDEN Investigator grant (project number 16599). Parts of this research were supported by the Australian Research Council Centre of Excellence for All Sky Astrophysics in 3 Dimensions (ASTRO 3D), through project number CE170100013.
D.O.J.\ is supported by a Gordon and Betty Moore Foundation postdoctoral fellowship at the University of California, Santa Cruz.  Support for this work was provided by NASA through the NASA Hubble Fellowship grant HF2-51462.001 awarded by the Space Telescope Science Institute, which is operated by the Association of Universities for Research in Astronomy, Inc., for NASA, under contract NAS5-26555.
The UCSC team is supported in part by NASA grant NNG17PX03C, NASA grants in support of {\it Hubble Space Telescope} programmes AR-14296 and GO-16239 through STScI, NSF grant AST-1815935, the Gordon \& Betty Moore Foundation, the Heising-Simons Foundation, and by a fellowship from the David and Lucile Packard Foundation to R.J.F.
J.H.\ was supported by a VILLUM FONDEN Investigator grant (project number 16599). W.J-G is supported by the National Science Foundation Graduate Research Fellowship Program under Grant No.~DGE-1842165. R.M. acknowledges support by the National Science Foundation under Award No. AST-1909796. R.M.\ acknowledges support by the National Science Foundation under Awards No. AST-1909796 and  AST-1944985. R.M.\ is a CIFAR Azrieli Global Scholar in the Gravity \& the Extreme Universe Program, 2019 and a Alfred P.\ Sloan Fellow in Physics, 2019. The Margutti team at Northwestern is partially funded by the Heising-Simons Foundation under grant \#2018-0911 (PI: Margutti).
E.R.-R.\ is supported by the  Heising-Simons Foundation, the Danish National Research Foundation (DNRF132) and NSF.
Based on observations obtained at the Gemini Observatory, which is operated by the Association of Universities for Research in Astronomy, Inc., under a cooperative agreement with the NSF on behalf of the Gemini partnership: the National Science Foundation (United States), National Research Council (Canada), CONICYT (Chile), Ministerio de Ciencia, Tecnolog\'{i}a e Innovaci\'{o}n Productiva (Argentina), Minist\'{e}rio da Ci\^{e}ncia, Tecnologia e Inova\c{c}\~{a}o (Brazil), and Korea Astronomy and Space Science Institute (Republic of Korea).  The Gemini-S/GSAOI observations in this paper were obtained under program GS-2020A-Q-221 (PI Kilpatrick).
Some of the data presented herein were obtained at the W. M. Keck Observatory, which is operated as a scientific partnership among the California Institute of Technology, the University of California and the National Aeronautics and Space Administration. The Observatory was made possible by the generous financial support of the W.\ M.\ Keck Foundation.
The authors wish to recognize and acknowledge the very significant cultural role and reverence that the summit of Maunakea has always had within the indigenous Hawaiian community.  We are most fortunate to have the opportunity to conduct observations from this mountain.
This work makes use of observations from the LCO network through program NOAO2019B-004 (PI Kilpatrick).
Based on observations made with the NASA/ESA {\it Hubble Space Telescope}, obtained from the data archive at the Space Telescope Science Institute. STScI is operated by the Association of Universities for Research in Astronomy, Inc. under NASA contract NAS 5-26555.
This work is based in part on observations made with the {\it Spitzer Space Telescope}, which was operated by the Jet Propulsion Laboratory, California Institute of Technology under a contract with NASA. (AST-1911206 and AST-1852393).

\textit{Facilities}: Gemini (GSAOI), {\it HST} (WFC3), Keck (LRIS), LCO (FLOYDS, Sinistro), {\it Spitzer} (IRAC), Swope (Direct/4K$\times$4K)

\section*{Data Availability}

The photometry in this article are presented in the article.  Imaging and spectroscopy data presented in this article are available upon request.  The {\it Hubble Space Telescope} and {\it Spitzer Space Telescope} data are publicly available and can be accessed from the Mikulski Archive for Space Telescopes (\url{https://archive.stsci.edu/hst/}) and Spitzer Heritage Archive (\url{http://sha.ipac.caltech.edu/applications/Spitzer/SHA/}), respectively.

\appendix

\begin{deluxetable}{lccc}\label{tab:photometry}
\tablewidth{0pt}
\tabletypesize{\scriptsize}
\tablecaption{SN~2019yvr Photometry}
\tablehead{
\colhead{MJD (Epoch)} & \colhead{Filter} & \colhead{Magnitude (Uncertainty)\tablenotemark{1}} & \colhead{Source}}
\startdata
58849.333 (-4.947) & $B$ &  16.781 (0.012) & Swope \\
58850.355 (-3.925) & $B$ &  16.691 (0.011) & Swope \\
58851.322 (-2.958) & $B$ &  16.577 (0.013) & Swope \\
58852.308 (-1.972) & $B$ &  16.606 (0.014) & Swope \\
58853.365 (-0.915) & $B$ &  16.570 (0.014) & Swope \\
58869.334 (15.054) & $B$ &  17.772 (0.022) & Swope \\
58871.311 (17.031) & $B$ &  17.970 (0.020) & Swope \\
58872.330 (18.050) & $B$ &  18.053 (0.023) & Swope \\
58874.339 (20.059) & $B$ &  18.267 (0.047) & Swope \\
58876.301 (22.021) & $B$ &  18.234 (0.018) & Swope \\
58849.331 (-4.949) & $V$ &  15.822 (0.009) & Swope \\
58850.354 (-3.926) & $V$ &  15.701 (0.008) & Swope \\
58851.321 (-2.959) & $V$ &  15.647 (0.009) & Swope \\
58852.307 (-1.973) & $V$ &  15.648 (0.009) & Swope \\
58853.364 (-0.916) & $V$ &  15.565 (0.009) & Swope \\
58869.332 (15.052) & $V$ &  16.302 (0.011) & Swope \\
58871.309 (17.029) & $V$ &  16.440 (0.010) & Swope \\
58872.329 (18.049) & $V$ &  16.476 (0.011) & Swope \\
58874.338 (20.058) & $V$ &  16.679 (0.024) & Swope \\
58876.298 (22.018) & $V$ &  16.689 (0.009) & Swope \\
58846.307 (-7.973) & $g$ &  16.590 (0.014) & LCO \\
58847.304 (-6.976) & $g$ &  16.440 (0.013) & LCO \\
58848.301 (-5.979) & $g$ &  16.317 (0.012) & LCO \\
58849.326 (-4.954) & $g$ &  16.293 (0.009) & Swope \\
58850.296 (-3.984) & $g$ &  16.147 (0.021) & LCO \\
58850.351 (-3.929) & $g$ &  16.170 (0.008) & Swope \\
58851.317 (-2.963) & $g$ &  16.123 (0.009) & Swope \\
58852.302 (-1.978) & $g$ &  16.050 (0.007) & Swope \\
58852.338 (-1.942) & $g$ &  16.022 (0.011) & LCO \\
58853.359 (-0.921) & $g$ &  16.036 (0.008) & Swope \\
58853.420 (-0.860) & $g$ &  15.946 (0.012) & LCO \\
58854.506 (0.226) & $g$ &  16.011 (0.016) & LCO \\
58855.506 (1.226) & $g$ &  15.982 (0.010) & LCO \\
58856.668 (2.388) & $g$ &  15.902 (0.023) & LCO \\
58857.672 (3.392) & $g$ &  16.024 (0.032) & LCO \\
58859.020 (4.740) & $g$ &  16.130 (0.014) & LCO \\
58861.015 (6.735) & $g$ &  16.170 (0.026) & LCO \\
58862.012 (7.732) & $g$ &  16.226 (0.021) & LCO \\
58866.340 (12.060) & $g$ &  16.736 (0.014) & LCO \\
58867.422 (13.142) & $g$ &  16.817 (0.012) & LCO \\
58869.328 (15.048) & $g$ &  17.019 (0.011) & Swope \\
58869.636 (15.356) & $g$ &  16.993 (0.014) & LCO \\
58871.304 (17.024) & $g$ &  17.130 (0.013) & Swope \\
58872.325 (18.045) & $g$ &  17.204 (0.011) & Swope \\
58872.329 (18.049) & $g$ &  17.098 (0.015) & LCO \\
58874.327 (20.047) & $g$ &  17.371 (0.009) & Swope \\
58875.307 (21.027) & $g$ &  17.230 (0.015) & LCO \\
58876.287 (22.007) & $g$ &  17.447 (0.008) & Swope \\
58876.338 (22.058) & $g$ &  17.319 (0.015) & LCO \\
58877.425 (23.145) & $g$ &  17.340 (0.016) & LCO \\
58846.309 (-7.971) & $i$ &  15.551 (0.010) & LCO \\
58847.306 (-6.974) & $i$ &  15.427 (0.009) & LCO \\
58848.303 (-5.977) & $i$ &  15.321 (0.009) & LCO \\
58849.300 (-4.980) & $i$ &  15.192 (0.008) & LCO \\
58849.325 (-4.955) & $i$ &  15.181 (0.007) & Swope \\
58850.298 (-3.982) & $i$ &  15.133 (0.008) & LCO \\
58850.350 (-3.930) & $i$ &  15.028 (0.007) & Swope \\
58851.316 (-2.964) & $i$ &  14.977 (0.007) & Swope \\
58851.340 (-2.940) & $i$ &  15.063 (0.008) & LCO \\
58852.301 (-1.979) & $i$ &  14.942 (0.007) & Swope \\
58852.340 (-1.940) & $i$ &  15.009 (0.008) & LCO \\
58853.358 (-0.922) & $i$ &  14.878 (0.006) & Swope \\
58853.422 (-0.858) & $i$ &  14.858 (0.009) & LCO \\
58854.508 (0.228) & $i$ &  14.880 (0.013) & LCO \\
58855.508 (1.228) & $i$ &  14.839 (0.007) & LCO \\
58856.670 (2.390) & $i$ &  14.726 (0.013) & LCO \\
58857.674 (3.394) & $i$ &  14.789 (0.014) & LCO \\
58859.023 (4.743) & $i$ &  14.763 (0.008) & LCO \\
58859.681 (5.401) & $i$ &  14.726 (0.012) & LCO \\
58861.017 (6.737) & $i$ &  14.740 (0.013) & LCO \\
58862.015 (7.735) & $i$ &  14.783 (0.012) & LCO \\
58863.012 (8.732) & $i$ &  14.799 (0.012) & LCO \\
58866.342 (12.062) & $i$ &  15.077 (0.008) & LCO \\
58867.425 (13.145) & $i$ &  15.089 (0.008) & LCO \\
58869.327 (15.047) & $i$ &  15.086 (0.007) & Swope \\
58869.639 (15.359) & $i$ &  15.216 (0.008) & LCO \\
58871.258 (16.978) & $i$ &  15.265 (0.009) & LCO \\
58871.303 (17.023) & $i$ &  15.145 (0.007) & Swope \\
58872.324 (18.044) & $i$ &  15.166 (0.006) & Swope \\
58872.332 (18.052) & $i$ &  15.304 (0.008) & LCO \\
58873.331 (19.051) & $i$ &  15.353 (0.084) & LCO \\
58874.323 (20.043) & $i$ &  15.327 (0.054) & Swope \\
58874.328 (20.048) & $i$ &  15.342 (0.018) & LCO \\
58875.309 (21.029) & $i$ &  15.393 (0.008) & LCO \\
58876.341 (22.061) & $i$ &  15.473 (0.008) & LCO \\
58877.427 (23.148) & $i$ &  15.434 (0.008) & LCO \\
58878.613 (24.333) & $i$ &  15.518 (0.010) & LCO \\
58882.099 (27.819) & $i$ &  15.562 (0.008) & LCO \\
58846.308 (-7.972) & $r$ &  15.798 (0.010) & LCO \\
58847.305 (-6.975) & $r$ &  15.638 (0.009) & LCO \\
58848.302 (-5.978) & $r$ &  15.562 (0.008) & LCO \\
58849.299 (-4.981) & $r$ &  15.501 (0.010) & LCO \\
58849.323 (-4.957) & $r$ &  15.419 (0.007) & Swope \\
58850.297 (-3.983) & $r$ &  15.370 (0.009) & LCO \\
58850.349 (-3.931) & $r$ &  15.302 (0.006) & Swope \\
58851.314 (-2.966) & $r$ &  15.256 (0.007) & Swope \\
58851.339 (-2.941) & $r$ &  15.293 (0.009) & LCO \\
58852.299 (-1.981) & $r$ &  15.227 (0.007) & Swope \\
58852.339 (-1.941) & $r$ &  15.250 (0.009) & LCO \\
58853.357 (-0.923) & $r$ &  15.141 (0.007) & Swope \\
58853.421 (-0.859) & $r$ &  15.153 (0.009) & LCO \\
58854.507 (0.227) & $r$ &  15.174 (0.014) & LCO \\
58855.507 (1.227) & $r$ &  15.149 (0.008) & LCO \\
58856.669 (2.389) & $r$ &  15.106 (0.013) & LCO \\
58857.673 (3.393) & $r$ &  15.078 (0.017) & LCO \\
58859.021 (4.741) & $r$ &  15.119 (0.012) & LCO \\
58859.674 (5.394) & $r$ &  15.102 (0.013) & LCO \\
58861.016 (6.736) & $r$ &  15.197 (0.019) & LCO \\
58862.014 (7.734) & $r$ &  15.200 (0.013) & LCO \\
58866.341 (12.061) & $r$ &  15.567 (0.009) & LCO \\
58867.423 (13.143) & $r$ &  15.601 (0.008) & LCO \\
58869.326 (15.046) & $r$ &  15.664 (0.007) & Swope \\
58869.638 (15.358) & $r$ &  15.746 (0.009) & LCO \\
58871.301 (17.021) & $r$ &  15.748 (0.007) & Swope \\
58872.323 (18.043) & $r$ &  15.781 (0.007) & Swope \\
58872.331 (18.051) & $r$ &  15.840 (0.009) & LCO \\
58873.329 (19.049) & $r$ &  15.958 (0.087) & LCO \\
58874.321 (20.041) & $r$ &  15.893 (0.009) & Swope \\
58875.308 (21.028) & $r$ &  15.941 (0.008) & LCO \\
58876.340 (22.060) & $r$ &  16.056 (0.009) & LCO \\
58877.426 (23.146) & $r$ &  16.036 (0.008) & LCO \\
58878.612 (24.332) & $r$ &  16.078 (0.010) & LCO \\
58849.327 (-4.953) & $u$ &  18.413 (0.031) & Swope \\
58850.352 (-3.928) & $u$ &  18.454 (0.032) & Swope \\
58850.352 (-3.928) & $u$ &  18.455 (0.032) & Swope \\
58851.319 (-2.961) & $u$ &  18.569 (0.032) & Swope \\
58852.303 (-1.977) & $u$ &  18.622 (0.027) & Swope \\
58853.360 (-0.920) & $u$ &  18.743 (0.032) & Swope \\
58869.329 (15.049) & $u$ &  19.946 (0.186) & Swope \\
58874.329 (20.049) & $u$ &  20.772 (0.268) & Swope \\
58876.290 (22.010) & $u$ &  20.672 (0.252) & Swope
\enddata
    \tablenotetext{1}{All magnitudes were calibrated using Pan-STARRS DR2 photometry standards following procedures described in \citet{Kilpatrick18:16cfr} and \autoref{sec:imaging}.  All $uBVgri$ magnitudes are on the AB system.  All epochs are reported relative to $V$-band maximum.}
\end{deluxetable}

\bibliography{progenitors}

\end{document}